\pdfoutput=1
\documentclass{JINST}
\usepackage{lineno}
\usepackage{graphicx}
\usepackage{epstopdf}
\usepackage[obeyspaces]{url}

\newenvironment{itemize*}%
 {\begin{itemize}%
    \setlength{\itemsep}{0pt}%
    \setlength{\parskip}{0pt}}%
  {\end{itemize}}

\newenvironment{enumerate*}%
 {\begin{enumerate}%
    \setlength{\itemsep}{0pt}%
    \setlength{\parskip}{0pt}}%
  {\end{enumerate}}

\title{The calibration system for the photomultiplier array of the SNO+ experiment}
\dedicated{\today}

 \author{R. Alves$^a$,
 S. Andringa$^b$, 
 S. Bradbury$^c$,
 J. Carvalho$^a$, 
 D. Chauhan$^b$$^d$,
 K. Clark$^e$$^k$\thanks{Present address: Dept. of Physics, University of Toronto, 60 St. George St., Toronto, Ontario M5S 1A7, Canada.},
 I. Coulter$^f$\thanks{Present address: Dept. of Physics and Astronomy, University of Pennsylvania, Philadelphia, Pennsylvania 19104-6396, USA},
 F. Descamps$^g$,
 E. Falk$^e$,
 L. Gurriana$^b$,
 C. Kraus$^d$,
 G. Lefeuvre$^e$\thanks{Present address: Micron Semiconductor Ltd, Lancing Business Park, Lancing BN15 8SJ, United Kingdom.},
 A. Maio$^b$$^h$$^i$, 
 J. Maneira$^b$$^h$\thanks{Corresponding author.}, 
 M. Mottram$^e$, 
 S. Peeters$^e$,
 J. Rose$^j$,
 L. Seabra$^b$, 
 J. Sinclair$^e$,
 P. Skensved$^k$,
 J. Waterfield$^e$,
 R. White$^e$,
 J.R. Wilson$^l$
 \\
\llap{$^a$}Laborat\'{o}rio de Instrumenta\c{c}\~{a}o e 
 F\'{\i}sica Experimental de Part\'{\i}culas and Departamento de F\'{\i}sica, Universidade de Coimbra, 3004-516 Coimbra, Portugal,\\
\llap{$^b$}Laborat\'{o}rio de Instrumenta\c{c}\~{a}o e 
 F\'{\i}sica Experimental de Part\'{\i}culas, Av. Elias Garcia, 14, $1^{\circ}$, 1000-149 Lisboa, Portugal,\\
\llap{$^c$}School of Physics and Astronomy, University of Leeds, Leeds LS2 9JT, United Kingdom,\\
\llap{$^d$}Dept. of Physics and Astronomy, Laurentian University, Sudbury, Ontario P3E 2C6, Canada,\\
\llap{$^e$}Dept. of Physics and Astronomy, University of Sussex, Falmer Campus, Brighton BN1 9QH, United Kingdom,\\
\llap{$^f$}Dept. of Physics, Oxford University, Denys Wilkinson Building, Keble Road, Oxford,  OX1 3RH, United Kingdom.\\
\llap{$^g$}Nuclear Science Division, Lawrence Berkeley National Laboratory, Berkeley, California 94720, USA,\\
 \llap{$^h$}Dep.to de F\'{\i}sica, Faculdade de Ci\^{e}ncias da Universidade de Lisboa, Campo Grande, 
 Edif\'{\i}cio C8, 1749-016 Lisboa, Portugal,\\
\llap{$^i$}Centro de F\'{\i}sica Nuclear da Universidade de Lisboa, Av. Prof. Gama Pinto, 2, 1649-003 Lisboa, Portugal,\\
\llap{$^j$}Dept. of Physics, University of Liverpool, Liverpool L69 7ZE, United Kingdom,\\
\llap{$^k$}Queen's University, Physics Dept., Kingston, Ontario K7L 3N6, Canada,\\
\llap{$^l$}School of Physics and Astronomy, Queen Mary, University of London, 327 Mile End Road, London, E1 4NS, United Kingdom.\\
E-mail: \email{maneira@lip.pt}
 }

\abstract{
A light injection system using LEDs and optical fibres was designed for the calibration and monitoring of the photomultiplier array of the SNO+ experiment at SNOLAB. Large volume, non-segmented, low-background detectors for rare event physics, such as the multi-purpose SNO+ experiment, need a calibration system that allow an accurate and regular measurement of the performance parameters of their photomultiplier arrays, while minimising the risk of radioactivity ingress. 
The design implemented for SNO+ uses a set of optical fibres to inject light pulses from external LEDs into the detector.
The design, fabrication and installation of this light injection system, as well as the first commissioning tests, are described in this paper. Monte Carlo simulations were compared with the commissioning test results, confirming that the system meets the performance requirements.
}

\keywords{SNO+; PMT; calibration; LED; Light injection}

\begin{document}

\section{Introduction}
\label{sec:introduction}

In SNO+~\cite{ref:snoplus_proc,ref:snoplus_proc2} liquid scintillator replaces the heavy water previously used in the SNO (Sudbury Neutrino Observatory) experiment~\cite{ref:SNOnim} with the primary aim of searching for neutrinoless double-beta decay, and measuring neutrinos from nuclear reactors, the Earth, the Sun and supernovae.

\begin{figure}[]
\begin{center}
\includegraphics[scale=0.3]{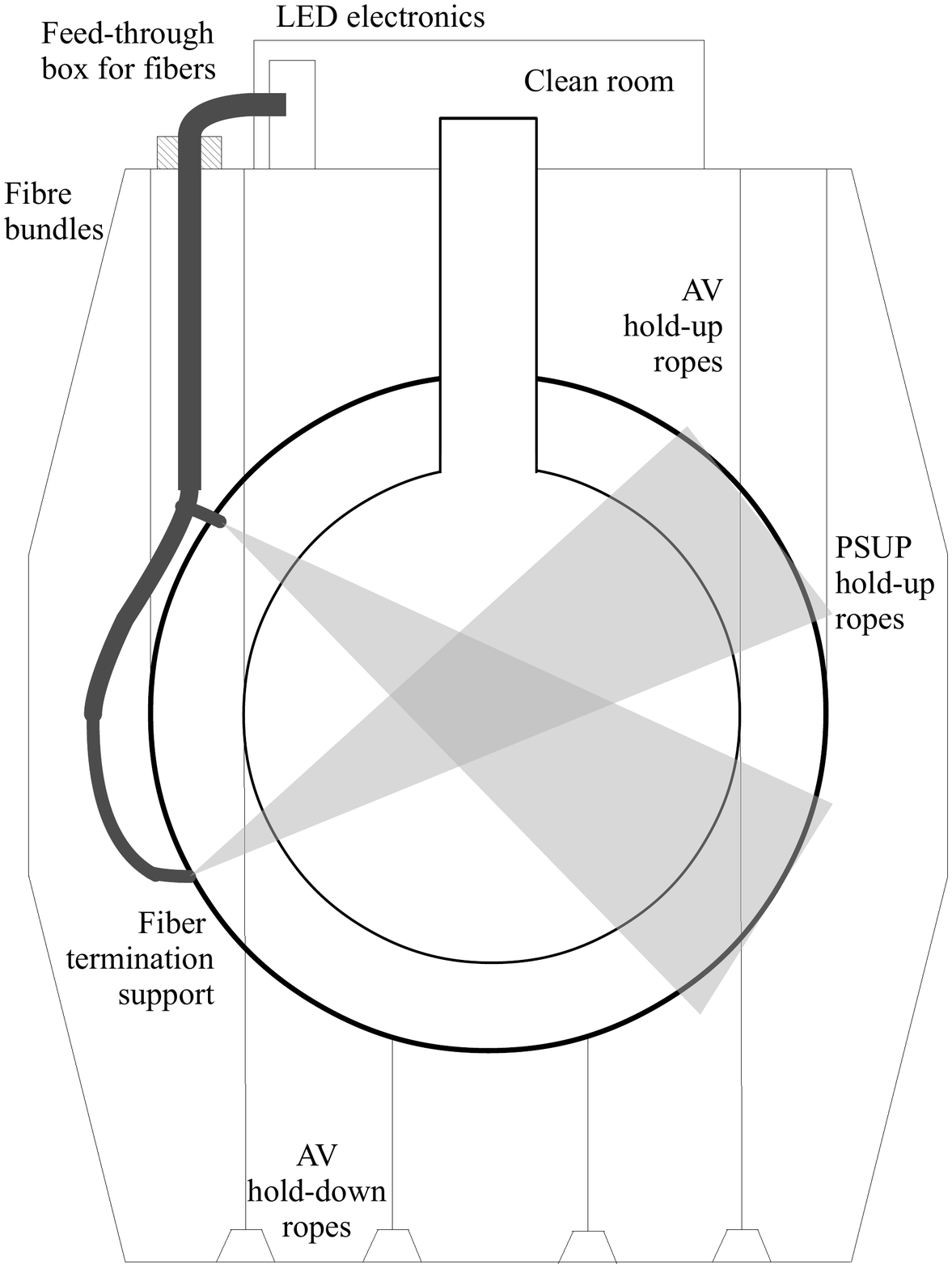}
\caption{Sketch of the SNO+ detector showing the calibration hardware on the deck above the acrylic vessel, as well as an example of the light injection points.}
\label{fig:snoplus_sketch}
\end{center}
\end{figure}

The central elements of the SNO+ detector are a 12\,meter diameter spherical acrylic vessel (AV) with a 5\,cm thick wall surrounded by an array of about 9500~photomultiplier tubes (PMTs) mounted in a 17.8\,meter diameter steel geodesic PMT support structure (PSUP). 
Each PMT is mounted inside a reflector to increase its optical coverage. 
The detector is located at a depth of 2092 meters, at SNOLAB, in the Vale Creighton Mine near Sudbury, Canada, providing effective shielding against cosmic ray muons and the neutron flux resulting from muon interactions. 
The inner-most detector volume is shielded from neutrons and gamma rays produced by natural radioactivity in the PMTs and surrounding rock, by about 7\,kilotonnes of ultra-pure water (better than ppt level for $^{238}$U and $^{232}$Th contamination), contained in a 34\,m-high cylindrical cavity covered with a radon-barrier. 
A system of field-compensation coils embedded in the cavity wall cancels the vertical component of the earth's magnetic field, such that the remaining effect is a 2.5\,\% reduction in photon detection efficiency\,\cite{ref:SNOnim}. No preferential direction of the first dynode was chosen when mounting the PMTs in their structure.
Detector electronics, as well as the calibration systems hardware, are located on the deck within the top part of the cavity. 
A sketch of the SNO+ experimental setup is shown in Figure~\ref{fig:snoplus_sketch}.
The liquid scintillator, Linear Alkylbenzene (LAB) with 2\,gL$^{-1}$ of PPO (2,5-diphenyloxazole), was chosen because of good optical properties -- high light yield and attenuation length --, a high flashpoint and its chemical compatibility with acrylic. 
With the new target material, several changes to the detector were needed, including new scintillator processing and purification systems, new trigger and readout electronics and new calibration systems. 
Structural improvements were required since the relative density of LAB is 0.86, causing a significant buoyant force on the scintillator-filled AV. 
This force is countered by means of a rope net\,\cite{ref:ropenet} that is anchored to the floor of the detector cavity.  

The use of liquid scintillator opens up the range of physics goals, making SNO+ a multi-purpose experiment. 
The main goal of SNO+ is the search for neutrinoless double-beta decay ($0\nu\beta\beta$), by loading a large mass of a suitable isotope into the liquid scintillator. 
When compared to experiments based on solid state detectors, the strategy of SNO+ is to compensate the lower energy resolution with a large mass of isotope and low background.
The chosen isotope for SNO+ is $^{130}$Te, with high natural abundance and favourable ($0\nu\beta\beta$) nuclear matrix elements and phase space, and a relatively small two-neutrino double seta decay rate. 
The optical absorption of Te in the loaded scintillator impacts the detector performance, limiting the initial concentration to $0.3\%$. 
Without the Tellurium loading, several solar neutrino measurements can be carried out at SNO+, including a precision measurement of the pep neutrino flux, observation of the low energy survival probability rise in $^{8}$B neutrinos and, possibly, direct measurements of the CNO neutrinos and the pp neutrinos. 
The observation of anti-neutrinos from nuclear reactors in Ontario and from the natural radioactivity chains of Uranium and Thorium present in the Earth's crust and mantle are additional goals. 
Throughout all the data-taking phases, the detector will also be part of the SNEWS\,\cite{ref:SNEWS} network monitoring for supernova neutrinos.

The physics goals of SNO+ require a low energy threshold, for measurements of solar neutrino elastic scattering signals, and for the tagging of decays of radioactive isotopes needed for the reduction of backgrounds in the $^{130}$Te $0\nu\beta\beta$ region-of-interest, close to the Q-value of 2.53\,MeV. 
The SNO+ threshold is effectively determined by the $^{14}$C background, at about 200\,keV. In this energy range, backgrounds from natural radioactivity are much higher than in the energy range of SNO (>\,3.5\,MeV). 
The requirements in terms of limiting the ingress of any radioactivity, especially $^{222}$Rn,  are consequently much tighter, which poses constraints to the operation of the detector calibration systems. In SNO, sources\,\cite{ref:laserball} were deployed inside the AV about once a month, but in SNO+ the requirement on the maximum amount of $^{222}$Rn acceptable in the scintillator cover gas system limits the planned deployment frequency to twice a year. 
The light injection system described in this paper was designed to obtain high quality, frequent calibration and monitoring data while avoiding the deployment of a source\,\footnote{The Borexino experiment\,\cite{ref:borexino}, that also employs a large volume of liquid scintillator for low background measurements,  previously developed a calibration system\,\cite{ref:borexino_fibres} with similar requirements for their array of 2200\,PMTs. }  
\footnote{In addition to optical sources, SNO+ will also use radioactive sources for calibration, as well as taking advantage of the spectra from radioactive isotopes present in the scintillator mixture and external detector materials.}.

This paper describes the PMT calibration system requirements, design considerations and characterisation measurements.
The overall requirements of the system are discussed in Section~\ref{sec:requirements}. 
The LEDs and their electronics pulse drivers are described in Section~\ref{sec:led}, followed by the results of the fibre characterisation tests in Section~\ref{sec:fibres}. The installation of such a system in an experiment like SNO+ is subject to several constraints, in terms of size and installation logistics, as well as cleanliness and radioactivity content. 
These design considerations are addressed in, respectively, Sections~\ref{sec:mechanical} and \ref{sec:mat_test}. Section~\ref{sec:control} describes the trigger and electronics control system. 
Finally, Section~\ref{sec:performance} describes the first commissioning tests with the full system chain, with data from air-filled detector runs, and the expected performance with scintillator, from simulation studies.

\section{PMT calibration system requirements}
\label{sec:requirements}

SNO+ is expected to detect between 200 and 500 photoelectrons per MeV from scintillation, depending on whether the scintillator is loaded with double-beta decay isotope and at what concentration.
At energies of 10\,MeV, above the range of most of the main physics goals, the expected average number of photoelectrons per PMT in SNO+ is such that the majority of PMT hits in SNO+ events will be single-photoelectron (SPE) hits, but with a significant fraction of hits by multiple photoelectrons.

The experimental observables in SNO+ will be the integrated charge of each PMT pulse, and the time at which that pulse crosses a given fixed discriminator level. 
In order to obtain a reconstructed position from the time measurement, it must first be corrected for any channel-dependent offsets, and for the discriminator time walk effect. 
In order to obtain a reconstructed energy from the charge measurement, it must be corrected for any gain variations of the PMTs, and also for the dependence of the light collection efficiency as a function of the event position. 
The measurement and monitoring of these three quantities -- PMT channel time offset, discriminator time walk, and gain -- is the main goal of the PMT calibration system. 

The position reconstruction is based on time-of-flight and as such, is directly linked to the time calibration of the entire detector. 
The time response depends on the decay time of the scintillator signal, on the SPE PMT resolution (dominated by the transit time spread), on the overall synchronisation of the PMT array, and on the correction of the time walk effect\cite{ref:cameron,ref:PMT_test}. 
The criterion for the accuracy of the synchronisation is to make the uncertainty in PMT time delays and walk corrections subdominatn with respect to the other effects -- scintillator decay time and PMT transit time spread -- in the reconstruction of the vertex position of events. 
These two effects can be measured by analysing the time-versus-charge distributions acquired with SPE pulses, since the width of the SNO+ PMT's charge distribution is large enough to populate all the relevant charge regions. 
Alternatively, they can be measured separately by using multi photo-electron (MPE) pulses, that do not suffer from the time-walk effect, to measure the channel-dependent delays. Once those are determined, SPE pulses can be used to constrain the time-versus-charge dependence independently of the individual channel delays.
Given the decay time of the scintillator signal, which is around 4.6\,ns (fast component) for the LAB/PPO mixture, and the SPE PMT resolution (dominated by the transit time spread), which is 1.7\,ns (standard deviation) for the SNO PMTs\,\cite{ref:SNOnim}, we set the PMT time calibration accuracy goal at 1\,ns.

In addition to the timing, the gain of the PMTs is an important parameter of the detector response, namely for energy reconstruction. 
Time and charge information are measured for each PMT pulse if it fires a discriminator with a fixed threshold, set for each channel at about $1/4$ of the average peak voltage for SPE pulses.
If the gain of the PMT changes, the photoelectron detection efficiency of that channel also changes. 
An additional goal of the calibration system is therefore to monitor the gain stability. 
From the expected photoelectron yield of the scintillation events, the required dynamic range for PMT gain monitoring is between 1 to 4 photoelectrons.

Other properties of the PMTs, such as after-pulsing, are also important for SNO+ since the precise tagging and measurement of coincidence events are needed for background rejection and antineutrino signal identification. This can also be investigated with a light emission system with flexible rate and intensity settings.

In addition to the PMT calibration goals, the efficiency of the data acquisition system to detect coincident signal bursts, such as those expected in the event of a supernova, can also be measured. For this purpose, the calibration system must allow a flexible program of pulse sequences from several channels.

In SNO, the PMT calibration was accomplished through the use of a laserball, a diffuse light source composed of a dye laser with selectable wavelengths and transmitting the light via optical fibres to a spherical diffuser that was deployed in the detector volume~\cite{ref:laserball}. 
The stringent radiopurity requirements for SNO+ forbid the regular immersion of external sources into the acrylic vessel: the system for calibration and monitoring of the PMT time response and gain is thus designed to be installed permanently outside the active volume of the detector.

The developed concept is that of an LED-based light injection system, with the light generation located on the deck above the detector to minimise the amount of potentially radioactive material close to the detector core and to allow access for maintenance of LEDs and electronics. 
The output intensity from each LED is measured with a dedicated PIN diode. 
The light then travels down a 47.75\,m fibre to the injection point, as shown in Figure \ref{fig:snoplus_sketch}.  
The fibre length consists of a 45\,m portion that connects the LED rack with all of the injection points while allowing for some slack on even the furthest mounts as well as a 2\,m patch cable used at the LED rack, and a 0.75\,m section inside each LED/driver box. All the fibres are constructed to be the same length in order to simplify the final timing calibration. 
Nevertheless, corrections due to different time-of-flight for PMTs located at different angles with respect to the axis of the light beam will be necessary. 
For most purposes, only one fibre is illuminated at each time, to avoid ambiguity about the origin of the detected light. 
A partial overlap of the beams is used to ensure consistency of the measurements with all fibres.  
Using more than one non-overlapping fibre at the same time can be useful, as a way to increase the data taking efficiency.

The characteristics and requirements for the different components of such a system are summarised below:
\begin{itemize}

\item The spectrum of the chosen LED must be in the wavelength region of the detected light for the several phases of the experiment and, within that range, should have a peak wavelength as high as possible in order to achieve the optimal direct transmission, since Rayleigh scattering of a medium decreases with increasing wavelength. 
Given the limiting factors -- the expected absorption of the Te-doped LAB-based scintillator, which is high below about 450\,nm, and the quantum efficiency of the PMTs, which drops quickly at higher wavelengths~\cite{ref:cameron} --  the optimal peak wavelength should be close to 500\,nm.
In order to monitor the PMT gain, the intensity of the LED light output should be measured with a precision of about  1\%.
The LED, its monitoring and its coupling to the optical fibre are described in Section~\ref{sec:led}.

\item A fast pulse LED driver is required. 
When coupled to the LED and the optical fibre, the driver must be capable of producing pulses with a FWHM of the order of 5\,ns to ensure that the measurement of the time offset of all PMTs can be done to a precision of 0.5\,ns within one day of low rate data taking.
In terms of intensity, the driver must provide pulses with a dynamic range in intensity such that the full system can provide SPE pulses in significantly different conditions in terms of optical attenuation of the light in the detector.
The driver should also have a wide range of frequencies, allowing a constant monitoring at a low rate of about 10\,Hz as well as fast dedicated calibration runs, in which rates up to about 10\,kHz are possible with the SNO+ read-out system.
The LED driver is described in Section~\ref{sec:led} as well.

\item The modal dispersion in the optical fibres must allow the FWHM of the transmitted LED pulses to be no higher than 5\,ns. 
In addition, the angular emission profile of the optical fibres must allow a sufficient illumination of all the PMTs even if a fibre fails. 
For the same redundancy reason, there should be two identical fibres per node. 
The fibre characterisation in terms of timing and angular distribution is detailed in Section~\ref{sec:fibres}, and the global coverage and performance studies are described in Section~\ref{sec:performance}.

\item Inside the detector, the mechanical mounting of the fibre ends on the PMT support structure must address two technical issues: the ease of installation and the material compatibility. 
Since most of the SNO detector hardware will not be replaced, the fibre system infrastructure must be adapted to the existing parts (Section~\ref{sec:mechanical}). 
In terms of materials compatibility, three criteria were checked for: total radioactivity of bulk material, emanation of Radon atoms, leaching when in contact with water(Section~\ref{sec:mat_test}).

\item The system control must fit in the main SNO+ trigger and acquisition framework while still retaining its designed flexibility: running quasi-continuous monitoring as well as dedicated calibration runs, monitoring the LED intensity as it flashes, flashing several LEDs in coincidence with different and adjustable intensities. 
Any time jitter in the trigger should not add to the observed rise time of the pulse. 
The calibration system control is described in Section~\ref{sec:control}.

\end{itemize}

In addition to PMT calibration, the experiment requires in-situ calibration of the optical properties of the detector media. 
The laserball source will be used for this task, but it will be essential to monitor the optical properties between source deployments. 
An extension of the LED-system, designed for the monitoring of the optical attenuation, is under development. 
This system relies on eight additional optical fibres with a narrower angular profile, that will be used with LEDs of different wavelengths. 
When complete, the attenuation monitoring system will be operated via the same driver and control circuitry as the PMT calibration system. 
The details of its requirements, design and performance will be described in a future paper.

\section{LED selection and driver circuit design}
\label{sec:led}

\subsection{LED selection and optical coupling}

\begin{figure}[t]
\centering
\includegraphics[width=0.7\textwidth]{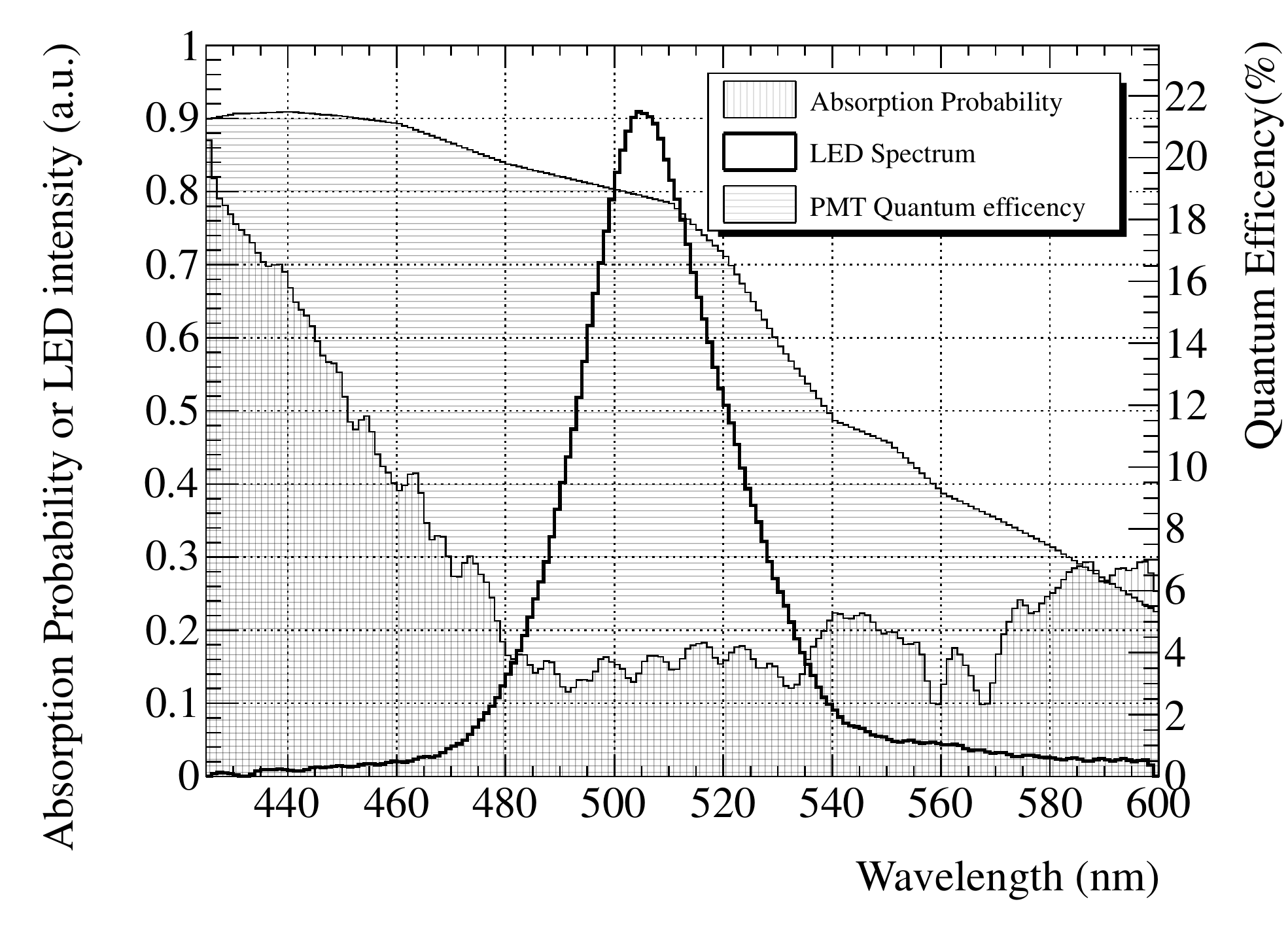}
\caption{\itshape {\small The absorption probability spectrum of Te doped LAB and PPO (vertical lines, left scale), the SNO+ PMT quantum efficiency (horizontal lines, right scale) and the average spectrum for all 96 LEDs (dotted area, left scale  in arbitrary units).}}
\label{fig:abs}
\end{figure}

The constraints on the LED emission spectrum are mainly dictated by the loaded scintillator absorption spectrum and the PMT quantum efficiency, as discussed in Section~\ref{sec:requirements}. 
These spectra, together with the measured wavelength distribution of the selected LEDs, that peaks at 503\,nm (see section~\ref{sec:ledqa}), are shown in Figure~\ref{fig:abs}.
The timing requirements for the LED pulses are derived from the timing characteristics of the PMTs. The single photoelectron transit time spread for SNO+ PMTs is, on average, 1.7\,ns\,\cite{ref:SNOnim}.
Since one of the goals of the calibration system is to determine various corrections -- inter-channel synchronisation and charge dependence --  to the timing measurement, the required precision for those corrections is set to $\sim$\,0.5\,ns, based on the criterion that these should not dominate over the transit time spread in their effect on the overall time precision.
The calibration corrections are determined from average parameters of the analysis over many calibration pulses, so in order to achieve a precision of about 1\,ns, LED pulses with widths of at most a few~ns are desirable.

To achieve this, we adopt a commonly used approximation for the low excitation LED modulation characteristics~\cite{ref:LED}. 
We describe the LED as an active region of volume $V_{a}$, driven by a constant current $I$ switched on at  $t=0$.  
Electrons are injected into $V_{a}$ increasing the carrier concentration, $n_{a}$.  
The emitted rate of photons per volume is $dn/dt=n_a/\tau_e$ where $\tau_e$  is the average time for a charge carrier to recombine and generate a photon. 
For a pn-junction current $I$, which is taken to be constant once the voltage exceeds the threshold, thus neglecting the large stationary charge densities either side of the LED's active region and their inherent capacitance,  the emitted rate of photons is then  

\begin{equation}
\label{eqn:rate}
\frac{dn}{dt} \left(t\right)=I \cdot \frac{1}{q_{e}\:V_{a}}\cdot \frac{\tau_l}{\tau_e}\cdot \left(1-e^{-\frac{t}{\tau_l}}\right).
\end{equation}

Here $\tau_l$ is the average time for a charge carrier to be lost from the active volume and $q_{e}$ is the charge of the electron. 
Equation~\ref{eqn:rate} implies that the current through the pn-junction is critical for the time to reach a desired photon emission rate. In practice this current is small compared to the much larger additional current required to increase the voltage for the substantial diode capacitance. 
In order to extinguish the LED on a time scale shorter than the charge carrier loss time $\tau_l$ a reverse current is needed to sweep out any remaining charges from the pn-junction. Details of the circuit used are presented in section \ref{sec:driver_design}.

An LED\,\cite{ref:LEDspecs} with a low resistance of 9\,$\Omega$ in comparison to the typical range of 40\,$\Omega$ to 100\,$\Omega$ was selected. 
This differential resistance is measured in the near linear region of forward current against forward voltage above the LED threshold voltage.  
The manufacturer was able to guarantee that all LEDs originate from the same production batch.

\begin{figure}[t]
        \centering
                \includegraphics[trim=0cm 0cm 0cm 2cm, clip=true, width=0.51\textwidth]{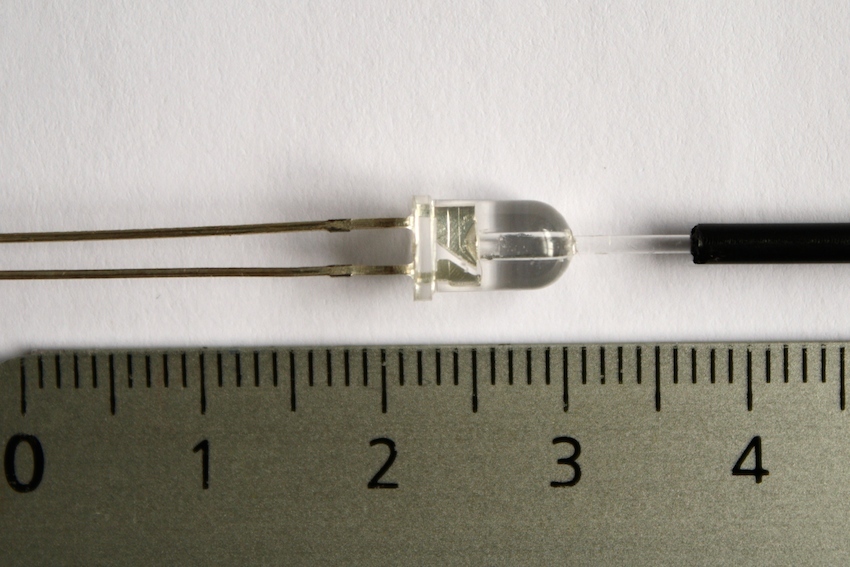}
                \includegraphics[trim=0cm 3cm 0cm 0cm, clip=true, width=0.47\textwidth]{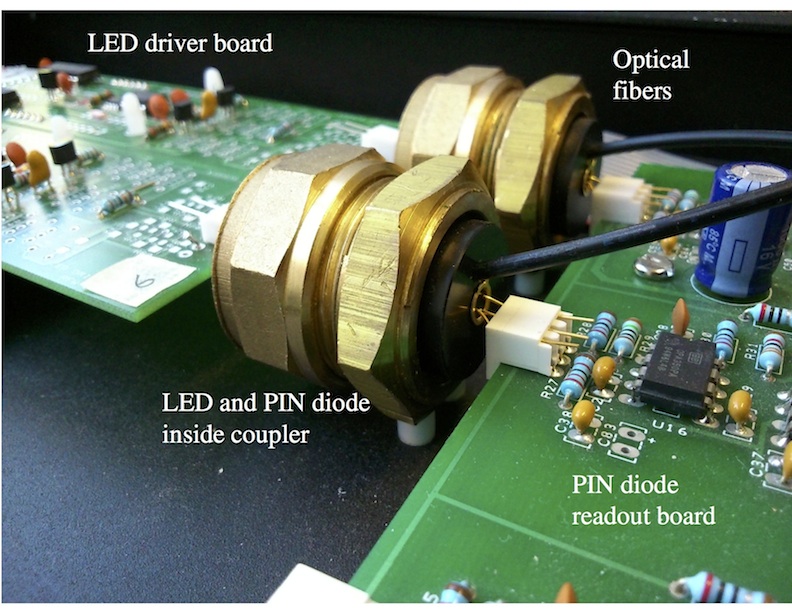}
 \caption{Left: Drilled LED with mounted optical fibre (ruler in cm). 
 Right:  Coupler containing LED and photodiode to monitor the number of photons at injection.}
\label{fig:LEDphotos}
\end{figure}

As we only require a small number of specifically selected LEDs, the coupling of the LED to the optical fibre was accomplished by drilling the LED lens to provide a socket in which an optical fibre of 1\,mm diameter was fixed, see Figure~\ref{fig:LEDphotos}. 
The fibre end was nominally 1\,mm from the diode.  
This method was shown to maximise the number of photons transmitted through the fibre while not altering the time profile of the pulse.

The LED and fibre are contained within a coupler, which also houses a photodiode mounted opposite the LED to monitor the average level of light injected into the fibre, see Figure~\ref{fig:LEDphotos}. 
The digitized output from this photodiode will be included in the data stream, to provide an estimate of the light intensity arriving at each PMT, that is independent from the LED driver settings.

\subsection{LED quality assurance}
\label{sec:ledqa}

The LED spectrum was measured for all 96 LEDs used in the PMT calibration system. Each LED was connected in series with a 550 $\Omega$ resistor, supplied with a constant 2.48\,V. 
The LEDs were connected to the spectrometer\,\cite{ref:spectrometer} via 47.5\,m of optical fibre, a 0.2\,m patch fibre was used to convert to the spectrometer's SMA (threaded type) connection. 
Each spectrum was recorded with an integration time of 40 ms and averaged over 20 scans. 
The spectral output was verified to be identical for pulsed and continuous operation.
Figure~\ref{fig:abs} presents the average wavelength spectrum over all 96 LEDs, in arbitrary units, showing a good match with both the expected scintillator absorption and PMT quantum efficiency. 
The mean peak wavelength is  $(505.6 \pm 2.6)$ nm.

The integral of the individual LED spectra provides an estimate of each LED output intensity.
The maximum difference between the response of any two LEDs is a factor of 18, and the typical spread (RMS/mean) is $43\%$. 
These LED response differences are large, and likely dictated by variations in the drilled fiber coupling rather than LED itself, but can be compensated by adjustments on the driver pulse intensity, as will be seen in section~\ref{sec:driverqa}.

\subsection{Driver circuit design and characterisation}
\label{sec:driver_design}

\begin{figure}[t]
\centering
\includegraphics[width=0.5\textwidth]{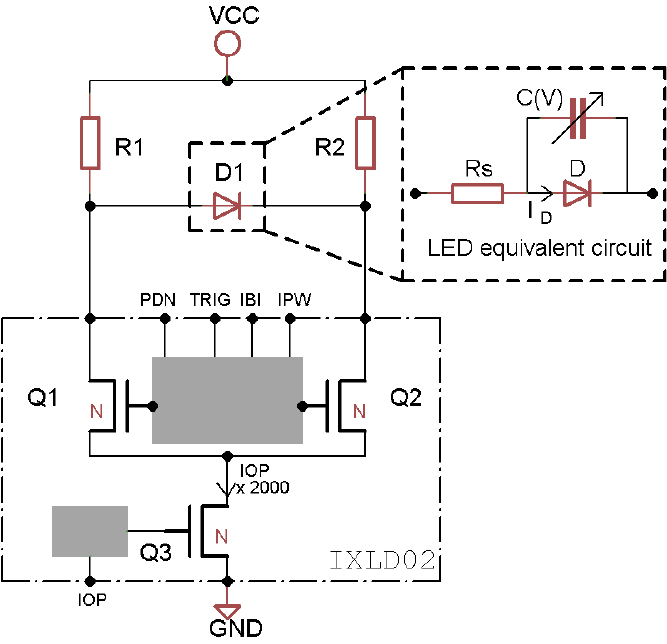}

\caption{ The LED driver circuit based on the IXLDO2SI commercial chip. Details are described in the text; here only the main components of the chip are shown, the rest of the circuit is represented by grey boxes. }
\label{fig:circuit}
\end{figure}

In this section we describe the driver circuit that is based on a commercial chip\,\cite{ref:circuit}, shown in Figure~\ref{fig:circuit}.

To create a small forward modulation current of a few mA through the pn-junction, shown as diode D in Figure~\ref{fig:circuit}, requires a much  larger total current through the LED series resistor into the capacitance C to then increase the voltage across the capacitor C to a value above the diode threshold voltage.  
For a maximum forward current of around 0.4\,A (series resistance R$_s$ = 9 $\Omega$, R1\,=R2\,=\,3.3\,$\Omega$, VCC\,=\,7\,V) and a capacitance of 200\,pF, the maximum slew rate, dV/dt\,=\,I / C, is 2\,V/ns. 
This is sufficient to create a short, one nanosecond period, bi-polar current pulse through the diode D, and thus create a short nanosecond time scale optical pulse.
The transistors Q1 and Q2, which are identical MOSFETs, are part of the integrated circuit, which also contains the gate drivers for Q1 and Q2 and the circuit to modify the operating current through the MOSFET Q3.
Initially transistor Q1 is open and transistor Q2 closed. 
To allow a forward voltage across the LED, Q1 closes and Q2 opens. 
Switching the LED off, returning Q1 and Q2 to their initial states creates a temporary reverse current, which  sweeps charge carriers from the LEDs active region. 
This significantly reduces the fall time of the LEDs pulse. 
Resistors R1 and R2 serve to limit the current through Q1 and Q2 as well as generating sufficiently large forward and reverse voltages across the LED. 
The choice of resistors is a compromise as large currents are needed for a high voltage slew rate, suggesting small resistor values, while the absolute transistor current limit and the need to generate forward and reverse diode voltages above threshold ask for large values.
The driver contains the MOSFET transistor drivers and the circuits needed to set the pulse width and amplitude, using currents labeled in Figure~\ref{fig:circuit} as IPW and IOP respectively. The operational current IOP activates the chip and sets the maximum bias applied to LED and in turn the maximum pulse amplitude; the pulse width current IPW sets the time that the voltages are inverted, i.e. the pulse width.

LED pulses are triggered via the TRIG input and between each pulse the current through Q1 or Q2 can be powered down via the PDN (power down) input. 
The  current flowing into the IBI pin acts as a baseline current with respect to the IPW current to compensate for internal delays\footnote{The naming conventions for input signals IOP, IPW, IBI, TRIG and PDN are from the circuit's manufacturer.}.

The circuit is always operated at maximum current by setting the control signal IOP to a maximum value.  
The electrical pulse width is then varied to change the number of photons generated per  pulse. 
Although the width of the optical pulse which is injected into the fibre varies, the optical pulse width of the light emerging from the fibre is in practice constant for a wide range of light intensities as shown in Table~\ref{tab:time_profile}. 
This occurs because the post injection increase in optical pulse width due to modal time dispersion in the fibre is larger than the width of the pulse generated by the LED.

Since the circuit is only active for $200\,\mathrm{ns}$ at each pulse, no heat sink is required.  
We have found that the circuit can be safely operated up to a rate of at least $10\,\mathrm{kHz}$, which is sufficient for the application. 
The circuit is operated at VCC\,=\,7\,volts to avoid significant MOSFET leakage currents. 

A single photon counting technique was used to precisely characterise the time profile of the LED pulse. 
The technique uses a coincidence setup between a trigger from the driver and a signal from a low background PMT\,\cite{ref:pmt} working in single photon mode, as described in~\cite{ref:singlePE}. 
Although this technique is the most accurate way to measure the time profile of the LED pulse, it is resource intensive and thus not a feasible method for characterising the entire system. 
This section focuses on the characterisation measurements carried out with one LED and driver combination, while the quality assurance of the full 96-channel system is described in Section~\ref{sec:driverqa}.

\begin{figure}[t]
\centering
\includegraphics[width=0.5\textwidth]{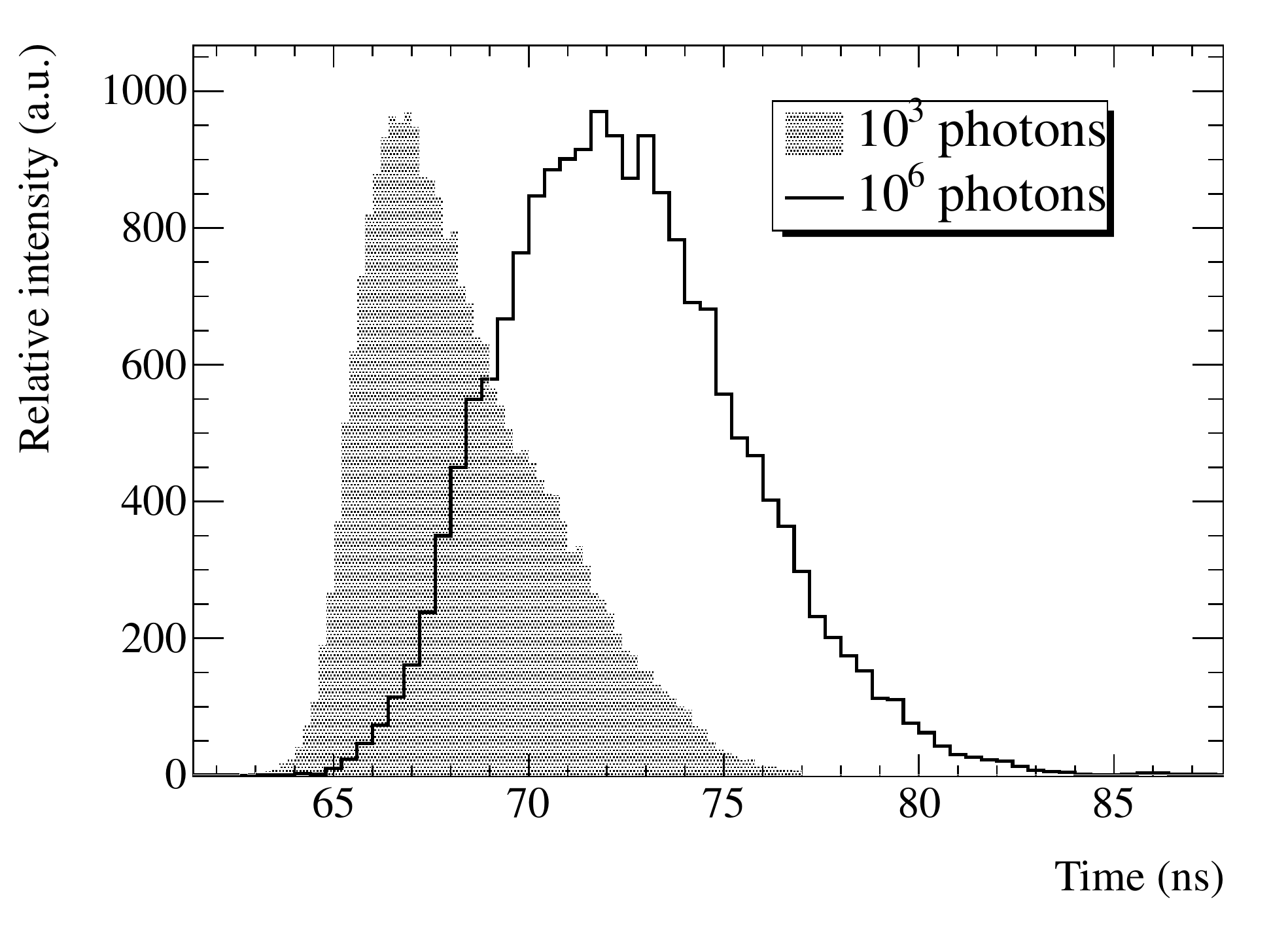}
\caption{Normalised time profiles from a single LED and driver combination, for pulses of $10^3$ and $10^6$ photons, in filled area and solid line, respectively. Measured using the single photon counting technique. }
 \label{fig:ten3ten6}
 \end{figure}

Table~\ref{tab:time_profile} shows a summary of the time profiles for four different amplitudes. 
The time profile for pulses of $10^3$ and $10^6$ photons is shown in Figure~\ref{fig:ten3ten6}, illustrating the broadening of the pulse at higher intensities. 
These measurements were made with the long fiber coiled in a bundle, ensuring mode-mixing. 
In the detector, the amount of fiber coiled will vary with their location in the detector, since all the fibers have the same length, but a minimum of 2\,m of excess is expected for the furthest ones. 
The coiling of the excess length is done close to the internal tip of the fiber, in order to improve the mode mixing and therefore the uniformity of the emitted light.

\begin{table}[t]
\centering
\begin{tabular}{| c | c | c | c |}
        \hline                        
        Number of Photons & Full Width (ns) & Rise Time (ns) & Fall Time (ns) \\ \hline \hline
       $10^{3}$ & $4.4$   & $1.6$ & $6.6$\\ \hline
        $10^{4}$ &$4.4$   & $2.0$ & $6.0$\\ \hline
        $10^{5}$ &$5.2$   & $2.0$ & $6.4$\\ \hline
        $10^{6}$ &$7.0$   & $4.2$ & $6.8$\\ \hline
    \end{tabular}
\caption{Summary of time profile measurements for a single LED and driver at varying intensities. 
All measurements made with the driver running at 1\,kHz. The measurement of the number of photons is performed after 45.5\,m of optical fibre, i.e., not including the 2\,m patch cord. 
The measurements of time profile of a single channel reported in this table are smaller, and more accurate, than the FHWM reported in Figure~\protect\ref{fig:fwhm_pho} due to systematic errors from the response time of the PMT and readout of the oscilloscope used in those measurements, that refer to all 96 channels.}
    \label{tab:time_profile} 
\end{table}

\subsection{Quality assurance of LED/driver channel}
\label{sec:driverqa}

All 96 driver and LED combinations used in the calibration system were characterised using the same 47.5\,m long optical fibre used in the SNO+ detector, coupled to the same PMT mentioned earlier\,\cite{ref:pmt} with the gain of the photomultiplier set by varying its control voltage between 0.6 V and 0.8 V depending on the pulse intensity. 
Instead of using the single photon counting technique, the  PMT signal was recorded with an oscilloscope\,\cite{ref:scope}. 
Mismatched impedance and the response times for the PMT and oscilloscope contributed to the broadening of time profile of the pulses; while undesirable, this was a systematic effect for all channels.

The PMT gain was characterised as a function of the applied voltage using a portable light-meter\,\cite{ref:powermeter}.  
The number of photons per LED pulse was obtained for two different LED light intensity settings by taking the average of five light meter readings, each of which was itself taken over two seconds with the LED pulsing at a known rate in the kHz range.  
This number of photons per pulse was then compared with integrated voltages from the PMT for the same LED and settings.  
These were taken at seven applied voltages between 0.56\,V and 0.85\,V, with the integration using the average of 1000 pulses for each reading.  
The relationships between the applied PMT voltage and the resultant gain were consistent between the two LED intensity settings.

The measured minimum photon output of each board/LED combination averaged at 232 and varied in the range of $20 - 510$ photons per pulse. 
The measured maximum photon output averaged at $1.13\times10^{6}$  and in the range of $0.45\times10^{6}  - 1.85\times10^{6}$ photons per pulse. 
These measurements showed that, even with the high variations in LED light output mentioned in Section~\ref{sec:ledqa}, after the LED/driver board pairing the uniformity allows  all channels to fall within the required operational range of $10^{3}  - 10^{5}$ photons per pulse.

\begin{figure}[t]
\centering
\includegraphics[width=0.7\textwidth]{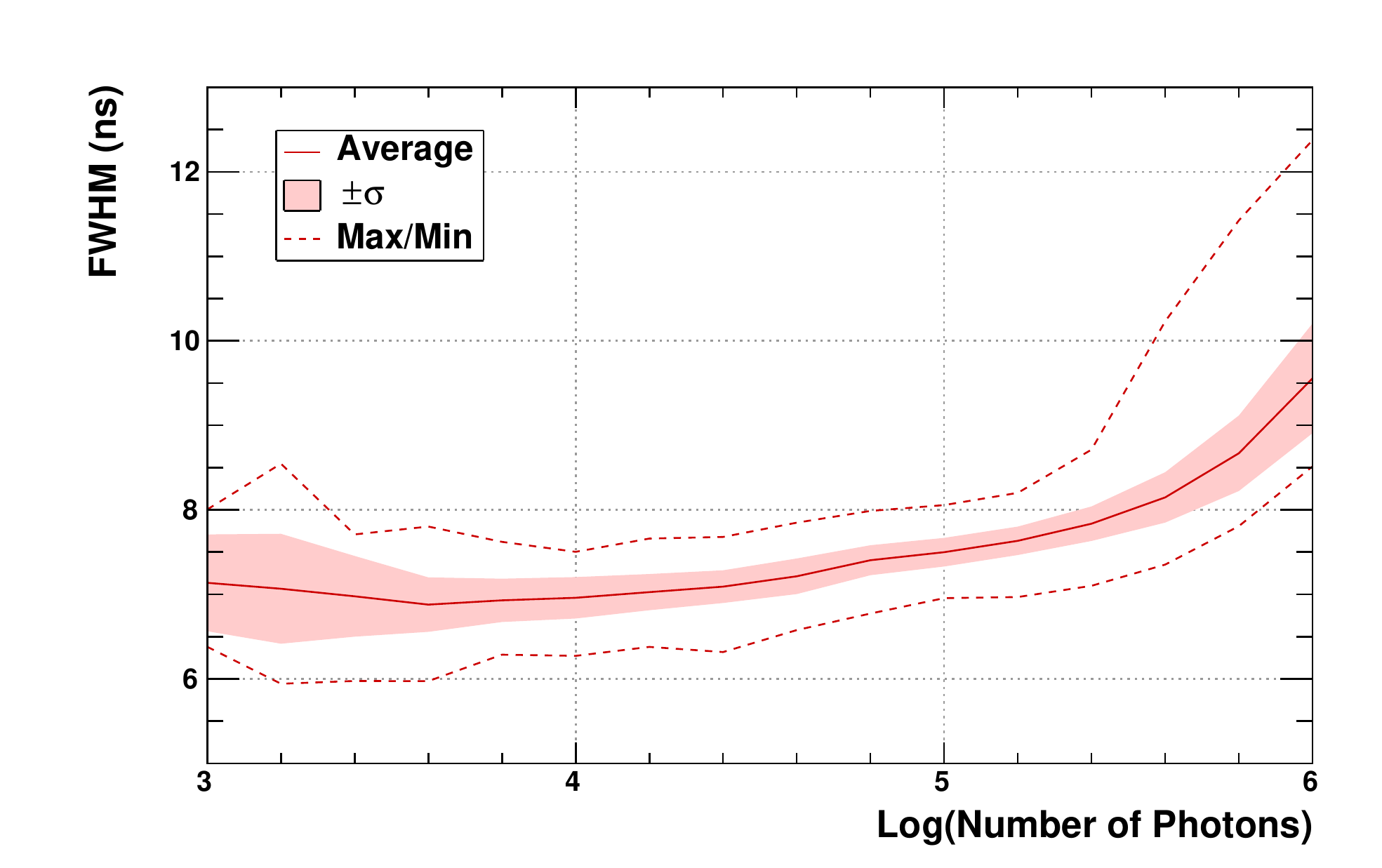}
\caption{Full Width Half Maximum (FWHM) of the measured light pulses from all the 96 LED and driver combinations as a function of their photon output. 
The light pulses are measures after going through 47.5 m of optical fibre. 
The FWHM reported in this figure is larger than the true FWHM reported in  Table~\protect\ref{tab:time_profile} due to systematic errors from the response time of the PMT and readout of the oscilloscope. 
The true full width half maximum is lower due to systematic errors from the response time of the PMT and readout of the oscilloscope. 
The LED driver was operated at maximum current while varying the electrical pulse width to control the number of photons per pulse. 
The observed increase in optical pulse width as the intensity reaches $10^{6}$ photons per pulse is due to the electrical pulse width for high intensities starts to exceed the modal time dispersion in the optical fibre.}
\label{fig:fwhm_pho}
\end{figure}

Figure \ref{fig:fwhm_pho} shows the time pulse full width at half maximum (FWHM) as a function of the number of photons. 
Due to the lower precision of the readout method, these time width readings are higher than those obtained by single photon counting (see Section~\ref{sec:driver_design}). 
Nevertheless, these measurements are useful since they were carried out for all 96 channels and show the response uniformity of the system as a whole. 

Before the quality assurance tests, the driver boards were subjected to a basic electronics test, and  3 out of 99 boards failed.
To pass the quality assurance tests the driver boards must be able to produce a minimum photon intensity of $10^{3}$ or less and a maximum photon output of $10^{5}$ photons or greater, with $10^{6}$ photons being desirable. 
All 96 boards met these intensity requirements with 27 boards having less than the desirable maximum intensity $10^{6}$ photons. 
Each driver board must also be able to achieve a measurement of the time offset of all PMTs to the required precision of 0.25 ns. 
These must be achieved with a run time less than 24 hours with the driver boards pulsing at the physics run rate of 10 Hz. This will correspond to a FWHM of 6 ns at photon intensity of $10^{3}$.
All driver boards met this requirement even though the systematic errors from the response time of the PMT and oscilloscope readout are not taken into account.

The final system has a temperature sensor between each of the driver boards. However, the temperature inside the electronics rack is not expected to vary much: the power produced by the driver is minimal and the system will be located on the SNO+ deck. The deck is above the cavity with water, which is kept at a constant temperature of 10$^\circ$C. From SNO, we expect the temperature on the deck to vary by less than 0.1~$^o$C. The light output produced the LEDs varies of the order of a couple of percent per degree of temperature change. The response of the PIN diode is much less affected by temperature, and has been shown to track the light accurately.

\section{Fibre characterisation}
\label{sec:fibres}

\subsection{Fibre selection}

The choice of the optical fibre cables is dictated by constraints of various aspects of this system, as described in Section~\ref{sec:requirements}. 
For ease of installation, the total number of fibres must be minimised while keeping sufficient overlap of the illuminated areas to allow for the synchronisation of all the PMTs. We use polymethyl methacrylate (PMMA) step-index fibres from Mitsubishi\,\cite{ref:fibre} with a wide aperture which meet these requirements. 

For the timing calibration itself it is essential that the fibres do not add significantly to the rise time of the light pulse. 
Due to the higher numerical aperture, the time dispersion associated with PMMA fibres is higher than that of quartz fibres, for instance, so that the aperture and timing requirements conflict.
However, as will be shown below, the time dispersion of the chosen fibres is still low enough to meet the SNO+ requirements
\footnote{According to typical manufacturer's specifications, graded-index fibers could provide a smaller time dispersion than the step-index fibers, but their cost is significantly higher so we did not pursue that option.}.

Each optical cable has two fibres, ensuring redundancy at each injection point without multiplying the number of cables to install and with better mechanical robustness than a single fibre cable. 
Their 1\,mm diameter and large numerical aperture (0.5) guarantee a wide illumination area. 
The LED-side ends are two pre-mounted ST (bayonet type) connectors, while the detector-side end is a double-fibre connector to be inserted in the mounting plate. 
For convenience all the fibres have the same length, leaving the propagation delay constant in the entire system. 

Finally, the construction materials, jacket and connectors included, must not compromise the required level of cleanliness and radiopurity in the ultra pure water surrounding the acrylic vessel. 
The tests made for this purpose are presented in Section~\ref{sec:mat_test}. 

In this section, we describe how the attenuation, aperture angle and timing distortions of the fibres were fully characterised. 
Key parameters were measured for quality control of 110 double fibre cables, in order to select the best ones and keep track of small variations within the system.

\subsection{Angular and timing distributions}

\begin{figure}[!htpb]
  \begin{center}
    \includegraphics[scale=0.50]{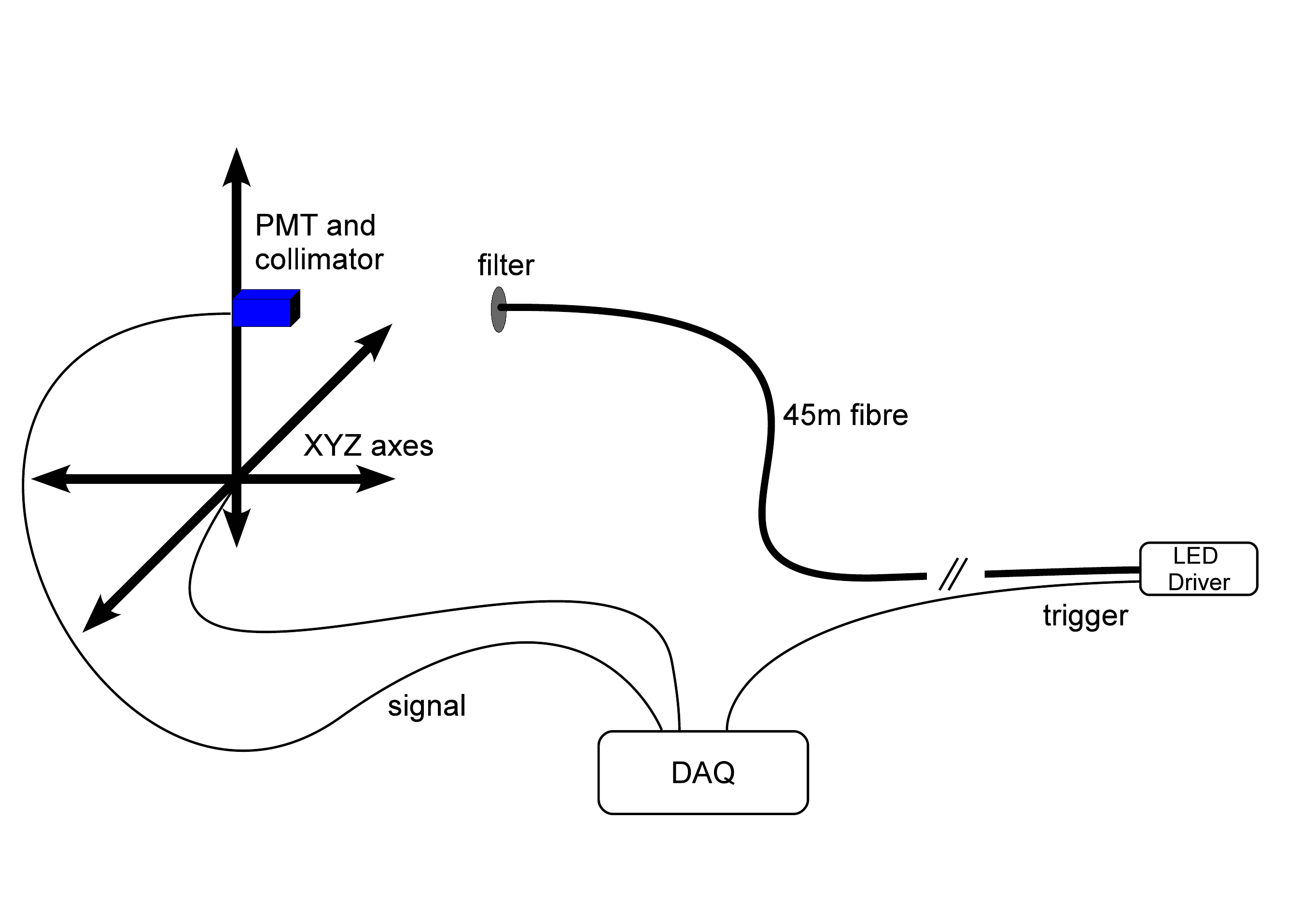}
    \caption{Diagram of the scan setup used to perform timing and aperture profiles. The PMT is fixed on an automated arm with three-dimensional movement, and its window blackened to limit the effective detecting area to the central 1\,mm$^{2}$. The double-fibre cable is fixed on a stand facing the PMT. A neutral density filter reduces its intensity so that the measurement can be operated in single photoelectron mode. In order to reduce backgrounds, the data acquisition requires a coincidence between the LED driver signal and the PMT output signal within a 50\,ns window.}
    \label{fig:2dsetup}
  \end{center}
\end{figure}

A specific setup was built to study the emission properties of the fibres, based on the previously mentioned PMT\cite{ref:pmt} moving in front of a fixed fibre and shown schematically in Figure~\ref{fig:2dsetup}. An LED driver similar to the final one (described in section~\ref{sec:led}) injects light into a patch cable, then into the full length  cable, thus reproducing the connections of the final system. The PMT has its window limited to a central 1\,mm$^2$ circular area. It is fixed onto an automated arm, whose three dimensional motions are independently actuated by a motion controller.  The PMT swipes an area in the plane perpendicular to the optical axis with a tuneable step and records the number of photons at each position using the trigger of the LED driver to feed a coincidence circuit.
In order to measure the true shape of the emitted light pulse it is sampled by a Time to Amplitude Converter (TAC) on a photon-by-photon basis. So the PMT works in SPE mode, made possible by the use of a very low LED intensity and a neutral density filter in front of the fibre connector.

\begin{figure}[t]
  \begin{center}
    \includegraphics[scale=0.50]{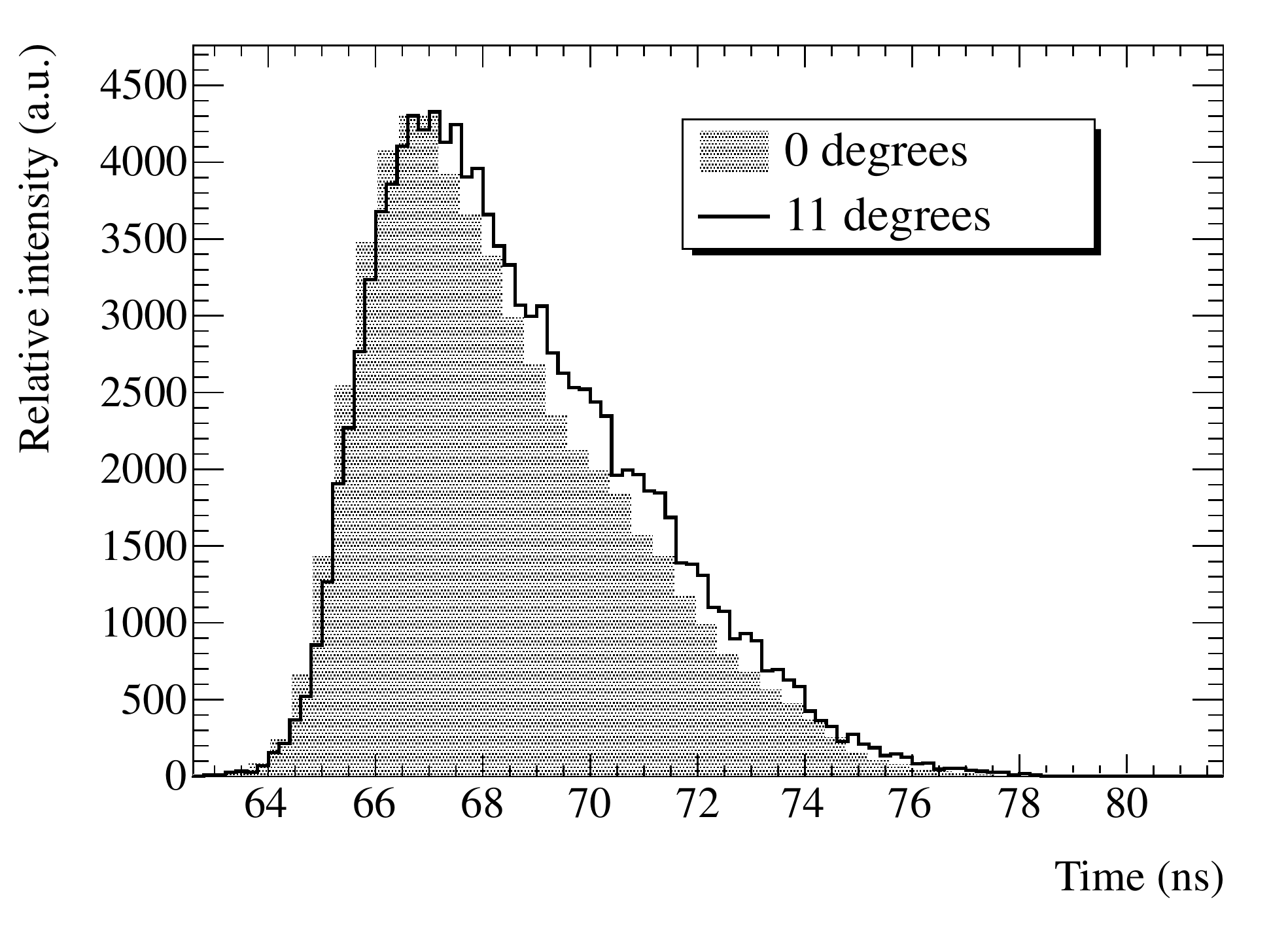}
    \caption{Time distribution of the light pulses emitted by the LED driver and going through the full length optical fibre, at the central position (filled area) and at an angle corresponding to 11$^{\circ}$ in water (solid line). The time offset is common to both histograms, and both are normalised to their integrals.}
    \label{fig:timingmeast}
  \end{center}
\end{figure}

The measurement was done at several positions, corresponding to different angles, in order to measure the spread of the pulse with respect to both angle and time. Figure~\ref{fig:timingmeast} shows these pulses at the central position and at an angle corresponding to 11$^{\circ}$ in water. The peaks are separated by 0.80$\pm$0.28\,ns and the widths at half maximum differ by 1.80$\pm$0.40\,ns. Such a dependence of timing on angle is expected for step-index fibres without  complete mode-mixing.
In order to achieve the calibration goal of approximately 0.5\,ns precision on the PMT timing, this effect will have to be corrected for in the calibration procedure. The measurement of the angular dependence of the fibre timing will be carried out in-situ, by comparing the timing offsets measured for the same PMTs with different fibres, and by comparing the offsets measured with the fibre system and the laser ball calibration source\cite{ref:laserball}.

\begin{figure}
  \begin{center}
    \includegraphics[scale=0.35]{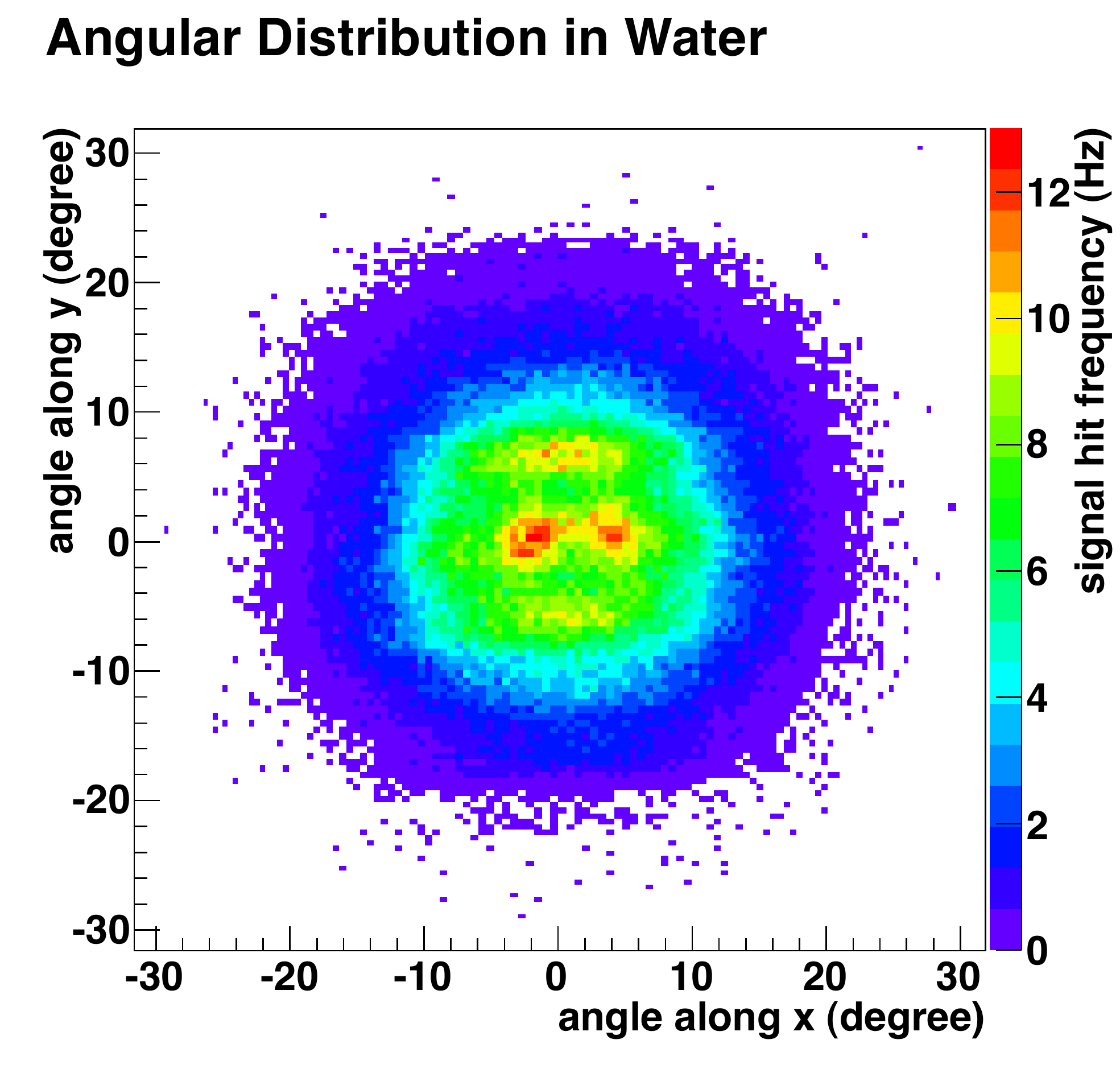}
    \caption{Two dimensional angular emission profile of a fibre. The white area corresponds to zero intensity. The steps in x and y are fixed during the measurement, hence the variable angular resolution, ranging from 0.6$^{\circ}$ at the center to 0.26$^{\circ}$ at the edges.}
    \label{fig:2dscan}
  \end{center}
\end{figure}

A wide two dimensional intensity scan was also performed. The 2D map of coincidences between PMT and LED driver is corrected for accidental coincidences, the geometry of the setup, and the refraction in and reflection on the filter.
 It is then converted into an angular intensity map in water, shown in Figure~\ref{fig:2dscan}. 
The fraction of the total light emitted within a cone of 14.5$^{\circ}$ from the center is found to be 80\%.

The bulk attenuation of the full length fibre is of the order of 3, but the effective dynamic range of the driver allows to make up for small differences that can arise.

\subsection{Fibre-to-fibre variation}

All the 110 double fibre cables were checked under the same conditions for timing, angular aperture and transmission. 
For the timing measurements, the PMT described above was connected to a 1\,GHz oscilloscope in order to record the rise and fall times of the signal and its total width, and the LED driver used was the same as in the previous section, which was similar, but not equal to the final version. 
Constant LED light sources at the same wavelength were used to measure the total light output for each fibre, first with a portable power-meter for a faster measurement to be repeated to check for possible damage during the installation, and then with a more precise measurement done in a dedicated set-up for optical fibre characterization.

\begin{figure}[t]
  \begin{center}
    \includegraphics[scale=0.60]{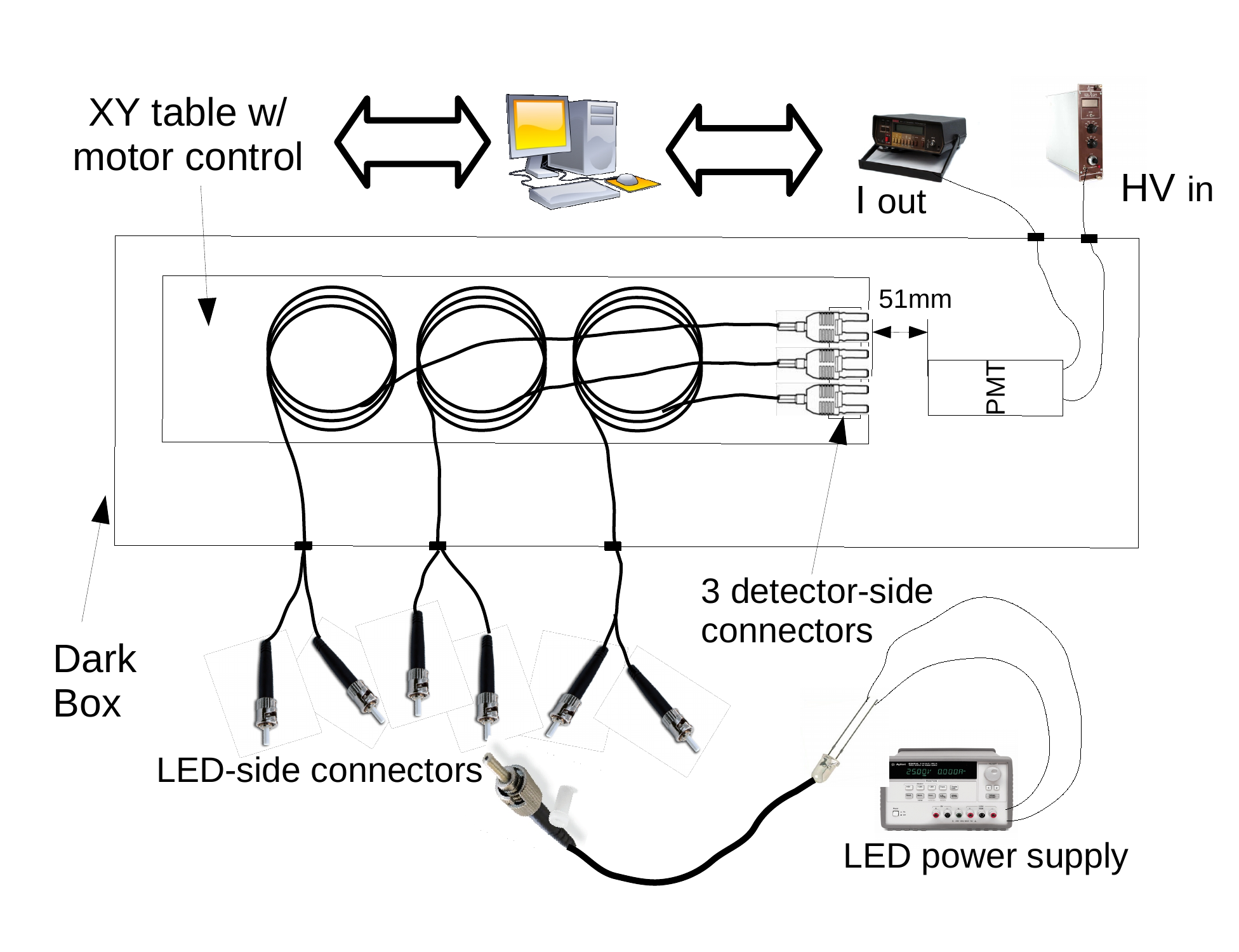}
    \caption{The interior of the dark box, showing the mounting of three 45\,m fibre cables in front of the PMT and light guide with the associated slit.}
    \label{fig:fibrometer}
  \end{center}
\end{figure}

The base automated test-bench described in~\cite{ref:fibrometer} was adapted to check the light output at different angles in a transverse scan, in order to cross-check the variation of the angular opening between fibres. 
The peak value and the integration of the angular scan also gives the variability in terms of light transmission. 
The set-up consists of a X-Y table in which one of the motors (X) was unused, while the other (Y) positioned the fibre at a given transverse position. 
The end of the fibre was 51\,mm from the detector, a PMT connected to a light guide with a Y-Z slit of 2.8 $\times$ 15\,mm$^2$. 
The Y-movement, in steps of 1\,mm, and the corresponding measurement of PMT current by a picoammeter, were computer controlled and fully automated.

\begin{figure}[t]
  \begin{center}
    \includegraphics[scale=0.50]{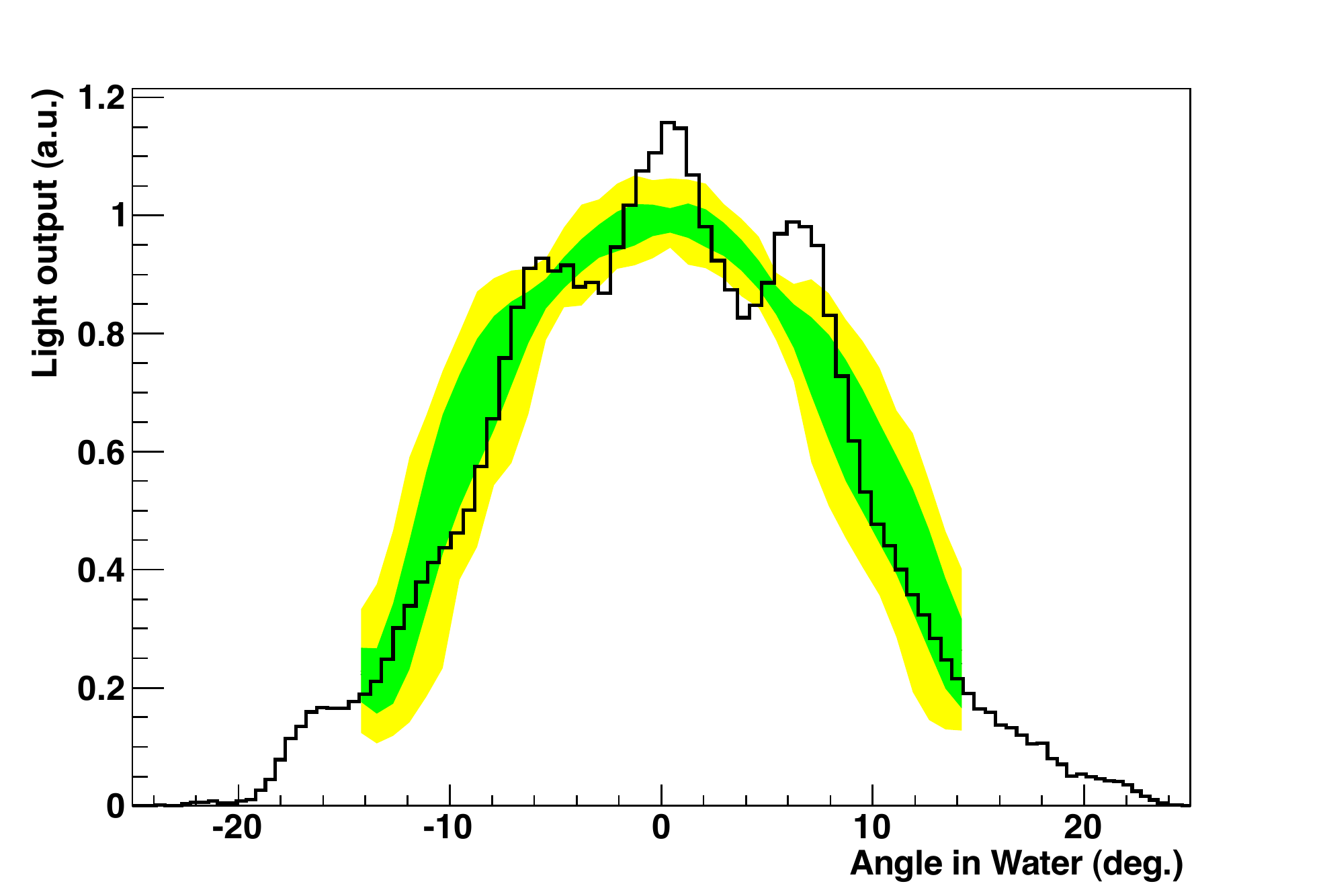}
    \caption{Comparison of independent measurements of the intensity profile of the fibre with respect to the angle in water: the black histogram is the profile of the detailed measurement with one fibre, from Figure~\protect\ref{fig:2dscan}; 
the coloured bands summarise the measurements of the full set of fibre cables. The yellow band shows the full spread and the green band represents the RMS, both over the total 220 fibres.}
    \label{fig:profilecomp}
  \end{center}
\end{figure}

Figure~\ref{fig:fibrometer} shows the interior of the dark box, adapted for this test. 
Three double-fibre cables can be mounted simultaneously in the XY-table. 
Since they are illuminated independently from outside the dark-box, they could all be checked in sequence without turning the PMT high voltage off,allowing for a reasonably fast check of all the 220 fibres. 
The light intensity was set to be above the system background which was measured and subtracted independently for each HV cycle. 
The edges of some of the angular distributions were cut for some of the fibres and showed some distortions, but the central part was consistent for all the six tested positions. 
The variations between the 220 fibres are shown in Figure~\ref{fig:profilecomp}, compared with the higher resolution test of a single fibre in the two-dimensional scan.

The shapes of the angular distribution and time signals were quite uniform, and consistent with the previous detailed studies.
The maximum intensity measured in the scan is well correlated with the power-meter measurements, even if it includes extra variation from the angular distribution shape. 
At installation, big drops in intensity will be used to identify damaged fibres so that they can be replaced. 
The figure-of-merit used to characterise the  angular distribution is the half width at 20\% intensity, after a correction to convert angles measured in air to angles measured in water (as will be the case in SNO+).

The selection of fibres was done in terms of the maximum intensity in the transverse scan, the angular opening (converted to water) at 20\% intensity, and the signal rise time, with values shown in Table \ref{tab:QA}. 
No fibre failed the angular criterion, although two were classified as preferably spares; in practice, even if the angular profile is slightly narrower than required, a larger intensity may allow the recovery of light at wider angle positions. 
Only one of the double fibre cables that had one fibre with lower transmission and the other with values close to the limit was excluded. 
In addition, four single fibres had timing or transmission values close to the established limits (some above, some below). 
Since their parameters were not far off from the requirements, and the other fibres in the same cable were well within the limits, these fibres were installed as well, but as spares. 

Two of the double cables (all with good parameters) were kept for ageing studies and future ex-situ tests. 

\begin{table}
  \begin{center}
    \begin{tabular}{|c|c|c|}
      \hline
      Characteristic & Mean and RMS & selection\\
      \hline
      $I_{max}$ (nA)   & 120.2 $\pm$ 21.9 & $>$60.0\\
      20\% Angle (deg.)&  14.5 $\pm$  0.2 & $>$13.5\\
      Rise Time (ns)   &   1.9 $\pm$  0.1 & $<$2.25\\
      \hline
    \end{tabular}
  \end{center}
  \caption{Fibre quality checking and selection was based on three parameters connected to transmission, numerical aperture and timing profile. The LED driver used was similar to the final version. } 
 \label{tab:QA}
\end{table}

\section{Mechanical structure}
\label{sec:mechanical}

The design of the mechanical structure of the PMT calibration fibres system is determined by the requirement of providing a coverage of the PMTs as uniform as possible, and by the access constraints to the existing detector structure. 

The 9500 PMTs are mounted with light reflectors in hexagon-shaped boxes called \emph{hex cells}, and made of ABS plastic (acrylonitrile butadiene styrene). 
These are assembled in flat panels, supported by a stainless steel geodesic structure (called PSUP). 

Most PMTs are mounted in triangular panels that follow the faces of the geodesic shape. Close to the geodesic nodes, 91 extra panels of haxagonal shape are mounted. 

The location of the hardware to support the calibration fibre terminations must not cause any shadowing on neighbouring PMTs nor suffer from light blocking by them. 
The sides of the \emph{hex cells} themselves provide a convenient mount point.
The \emph{hex cell} panels in the geodesic nodes of the PSUP were chosen for the location of the mount points, since they have a slightly wider gap with respect to the neighbouring panels, facilitating installation.
In addition, their distribution in the detector guarantees in a simple way a uniform coverage of all the PMTs, when all 91 nodes are used.

The number of LED-driver channels is 96, chosen to be slightly higher than the required 92 in order to ensure some redundancy in case of channel failure.
The number of fibres installed is also higher, 110 double cables, in order to have enough spares in case of breakage during installation. 
In addition to the 110 fibres of the PMT calibration system, SNO+ has an additional 25 single fibre cables for an additional optical calibration system that is not discussed here.

The mechanical design of the system needs to deliver the set of 135 fibre cables (including fibre cables for other systems) from the dry deck area, where the LED rack is located, into the detector, routing them around the PSUP and mounting their tip in the designated location on the hexagonal PMT cell in each of the 91 nodes.

\subsection{From the deck area to the detector}
The main constraint of this part of the design is the limited space available for feeding the 135 fibre cables into the detector area, since an access port with dimensions $46\,\times\,13$~cm is used. 
During the initial part of the installation, the fibres are left hanging from the deck down to the cavity floor in air, so they need to be able to support their own weight. 
Gas tightness is not a requirement for the feed-through, since the air volume is in contact only with the outer water volume, but light tightness is needed, to avoid light leaks into the detector.

\begin{figure}
\begin{centering}
\includegraphics[width=2.7in]{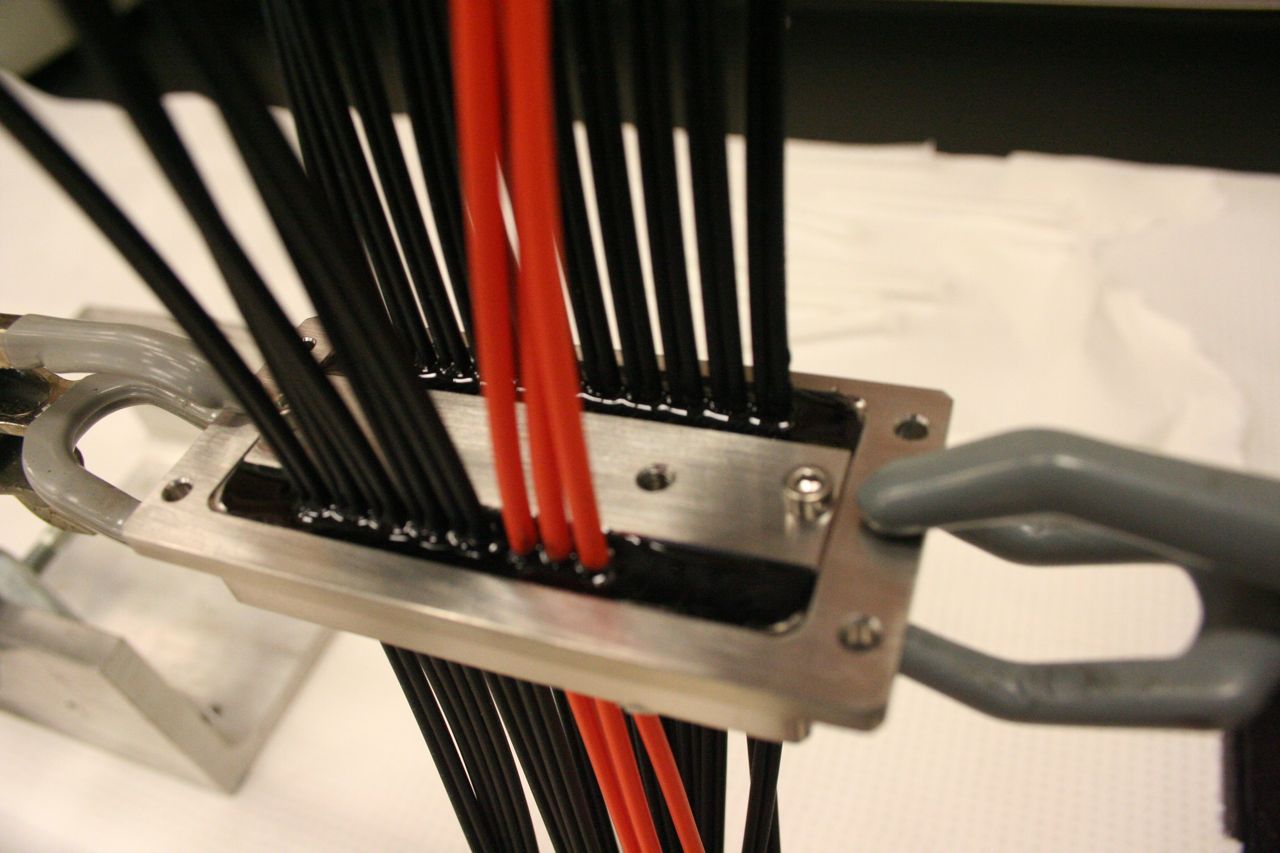}
\includegraphics[width=2.5in]{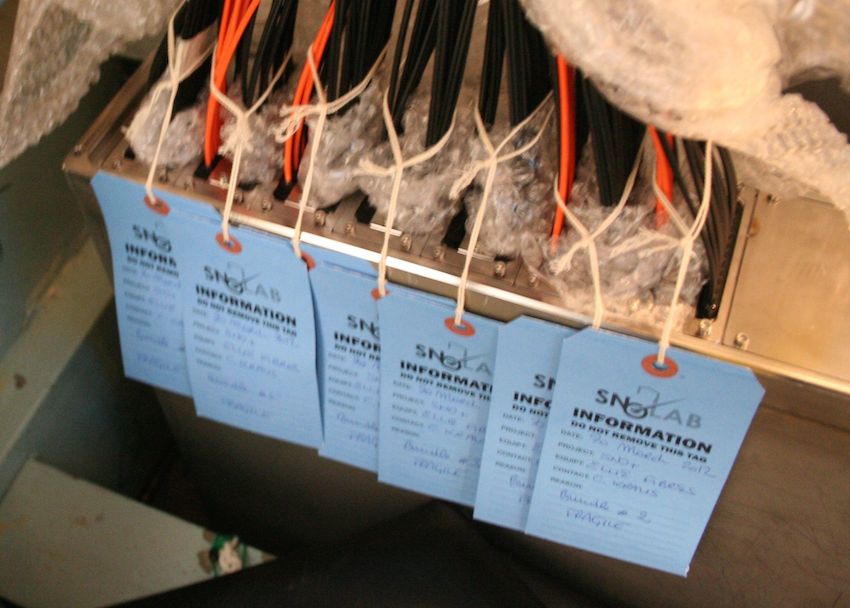}
\par\end{centering}
\caption{Deck feedthrough for all SNO+ calibration fibres. 
The left photo shows one of the eight feedthroughs, that can hold up to 17 double PMMA fibre cables -- 11 in the back row and six in front -- and six single quartz fibres -- in the front row, right. 
These feedthroughs are assembled in sequence on a raised flange, shown in the right photo, mounted on one of the spare PMT cable flanges in the deck area.}
\label{fig:feed-through}
\end{figure}

The chosen design consists of a set of eight feed-throughs, mounted on a raised flange, both shown in Figure~\ref{fig:feed-through}. 
Each feed-through is made of an inner and an outer piece with matching grooves, that hold up to 17 double and 6 single fibre cables. 
The length (12 mm) and diameter of the grooves is such that the compression on the fibre jacket once both pieces are fixed is enough to hold the fibre cables firmly in place. 
Nonetheless, in order to further secure them, and to improve the light tightness, an epoxy compound is applied to the feed-through. 

This modular design allows the application and curing of the compound to be done in the surface labs, reducing the number of installation steps to be done underground. 
Each bundle is brought underground with the respective feed-through in place, ready to be mounted in the flange. 
The feed-throughs can enter the flange from below, simply by rotating them 90 degrees with respect to their final position. 
Since the access port where the box is mounted is lower with respect to the deck area floor level, the feed-through flange is raised 50 cm with respect to the bottom flange. 
A rubber gasket seals the coupling between the bottom flange and the hole edge. 
To avoid corrosion from the humid air below, the whole box and feed-throughs are made of stainless steel.

From the feed-through box, the fibres hang straight down into the detector, with no bends until they reach the PSUP. 
Furthermore, to protect against accidental breaking, the selected fibre cable has a large tensile strength of 140 N, about 35 times the weight of a single cable. 
So even in the improbable case that during the installation a single fibre would hold the weight of a full bundle, the force would still be well below the fibre's tensile strength, since the largest bundle has 21 cables.

\subsection{Fibre tip mount points at the detector}

\begin{figure}
\begin{centering}
\includegraphics[width=3.5in]{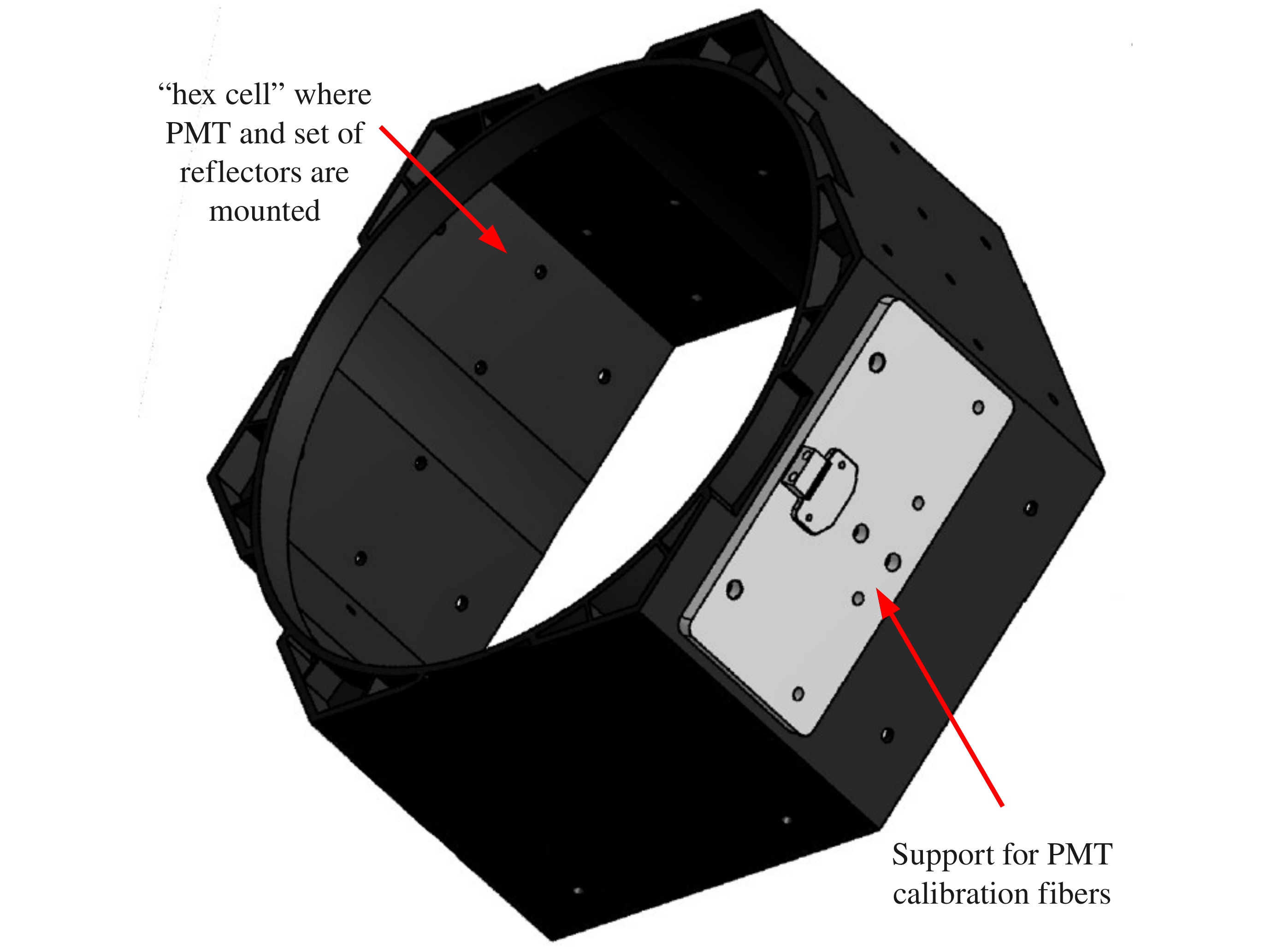}
\par\end{centering}
\caption{Mount points for the fibre cables.}
\label{fig:mount}
\end{figure}

The fibre tips are mounted in plates made from PETG (polyethylene terepthalate glycol) and designed to be attached to the PMT \emph{hex cells}. 
Since the available space between can be as small as 4-5 cm, and their inside part is not accessible, the fibre mount plates, shown in Figure~\ref{fig:mount}, are secured to the pre-existing holes of the \emph{hex cell} sides with nylon push-in rivets.

The fibre mount is designed for the latching fibre connector to simply clip in, for ease of installation. 
The two smaller holes visible in the center of the plate are used to secure cable ties that hold the fibre straight. 
The remaining holes are used to fix the plate with the rivets, and the four holes aligned in the lower edge also serve to secure a black plastic sheet held between the PMT panels, that serves as a light barrier between the inner and outer water volumes.

\subsection{Routing and installation sequence}

The routing of the fibres from the deck feed-through to the mount points in the PSUP is strongly dependent on the detector access method during the installation period. 
The scaffold or lift access available from the cavity floor reaches only the bottom mount points, about one third of the total. 
Manpower access from the deck to the top of the PSUP is possible and will be employed to install the top one sixth of the mount points. 
The remaining positions, about half of the total, will be accessed from a boat during the water fill.

The fibre bundles were initially brought into the cavity floor in spools, and the LED-side ends of the bundles were unwound and pulled up to the deck through the hole. 
The detector-side ends of the fibres were removed from the spools and routed to their mounts points, along the steel beams of the PSUP, by using a movable lift and a scaffold. 
For the installation during water fill, the remaining fibre bundles are grouped together based on the vertical level of the mount point, and a horizontal routing path will be followed. 
The length of the fibres is determined by the routing path of the mount points that are furthest away from the deck port. In order to minimise this length, the five furthest points furthest are routed with an alternate, shorter path.

For each fibre cable, a final quality control was carried out just prior to securing the fibre in its support, by shining continuous LED light on the LED-side end, and checking the transmitted intensity, using the same optical power meter used in the pre-installation quality control (see Section~\ref{sec:fibres}). 
If a significant drop is observed, the fibre is replaced by one of the spares, that are all grouped in one bundle. 
At the time of writing, one third of the fibres were fully installed, but all the fibres were brought into the detector and then tested. 
Only one fiber was found broken, and the second fibre in the same cable is not damaged.
The spare fibres that are not needed are installed in extra positions, in order to increase the redundancy of the light coverage.

\section{Materials testing}
\label{sec:mat_test}

The selection of materials and suppliers for all the system parts located within the detector obeyed different criteria on total bulk radioactivity, radon emanation, and compatibility with ultra-pure water. The tests that were carried out  and their results are described in the next sections.

\subsection {Radioactivity}

To determine the appropriate acceptance limits for contamination by natural radioactive isotopes (from the $^{238}$U, $^{235}$U and $^{232}$Th chains, and from $^{40}$K) in new parts to be installed in the SNO+ detector, the known amounts of contamination of the main detector parts of the SNO detector were used as a benchmark. 
The structure of the SNO+ detector is one of concentric shells of increasing radiopurity, each layer acting as a barrier for gamma radiation and  neutrons from radioactive decays originating in the outer shells. 
Therefore, the requirements on the new system parts in terms of contamination by these isotopes depend on the actual location where the specific part is to be installed.
Since the system will be installed in the PMT support structure (PSUP), close to the PMT array, with the fibre bundles coming out of the detector similarly to the PMT cables (see next section), the comparison parameter was chosen to be the total contamination in the array of PMTs and its associated cabling.

Simulations of the backgrounds induced by the PMTs in the SNO+ analysis windows for its main physics goals -- $^{130}$Te neutrinoless double-beta decay and solar neutrinos -- were carried out by the SNO+ collaboration\,\cite{ref:snoplus_proc}. 
These show that, while external backgrounds are responsible for limiting the fiducial volume for the planned measurements, the contribution from the PMTs is subdominant with respect to that of other components located closer to the scintillator volume.
Based on these considerations, the acceptance limit is set at a total amount of radioactive isotopes equivalent to 10\% of the PMT array and cables.

The materials used in each of the PMTs include approximately 100\,$\mu$g of $^{238}$U and $^{232}$Th~\cite{ref:SNOnim}. 
The total number of PMTs in SNO+ was assumed to be 9500 (installation and repairs are still ongoing so the final number is not precisely known at the time of writing), giving a total amount of each radioactive isotope of 0.95\,g, as shown in Table~\ref{tab:full_isotope_cont}. 
The levels of $^{235}$U and $^{40}$K in the PMTs are not included in this study.

The contamination measurements were done by low-background gamma counting with a HPGe detector at SNOLAB~\cite{ref:gamma_assay}. 
The results are available in reference~\cite{ref:counting_results} and the values shown in Table~\ref{tab:full_isotope_cont} are for a total of  110 full-length fibres, which includes the spares. 
As can be seen, the total amount of radioactivity introduced by the calibration system is well below the limit of 10\% of the PMTs plus cables.

\begin{table}[t]
\begin{center}
	\begin{tabular} {| c | c | c | c | c |} \hline
	Item & $^{238}$U (g) & $^{235}$U (g) & $^{232}$Th (g)  & $^{40}$K (g)\\ \hline
	
	Fibre cables & 3.83$\times$10$^{-4}$ & <1.33$\times$10$^{-7}$ & 6.81$\times$10$^{-6}$ & 6.13$\times$10$^{-1}$ \\
	Mounting plate & 4.61$\times$10$^{-5}$ & 2.12$\times$10$^{-7}$ & 3.67$\times$10$^{-6}$ & 1.10$\times$10$^{-2}$ \\
	Mounting parts & 3.79$\times$10$^{-4}$ & 2.44$\times$10$^{-7}$ & 2.36$\times$10$^{-5}$ & 6.61$\times$10$^{-2}$ \\ \hline\
	Total & 7.36$\times$10$^{-4}$ & 5.38$\times$10$^{-7}$ & 3.10$\times$10$^{-5}$ & 6.28$\times$10$^{-1}$ \\ \hline \hline
	
	\textit{PMTs} & \textit{9.50$\times$10$^{-1}$} & - & \textit{9.50$\times$10$^{-1}$} & - \\ 
	\textit{PMT Cables} & \textit{2.72$\times$10$^{-1}$} & \textit{2.72$\times$10$^{-1}$} & \textit{2.72$\times$10$^{-1}$} & \textit{3.6$\times$10$^{3}$} \\ 	\hline

\hline
	\end{tabular}
\end{center}
\caption{Full installation values for the amount of each isotope included in the calibration systems. The amount shown corresponds to 110 fibre cables and 9500 PMTs and PMT cables.}
\label{tab:full_isotope_cont}
\end{table}

\subsection{Radon emanation}

Emanation of $^{222}$Rn from the materials' surfaces must be checked independently from the bulk radioactivity due to the mobility of the Radon gas. 
The guidelines for acceptable levels were established by the SNO Collaboration~\cite{ref:sno_radon} at 14\,mBq/m$^{3}$ for the outer H$_2$O region, where the fibres will be located. 
The actual measured value of Radon contamination in that region was 4.3\,mBq/m$^{3}$. 
We take that value as a guideline and, similarly to the bulk radioactivity, require the additional activity introduced by the calibration system to be less than $10\%$ of the guideline value.

A sample of the fibre cable was measured at Queen's University. 
The measurement setup is described in~\cite{ref:radon_queens} and the results are shown in Table~\ref{tab:rad_results}. In addition to presenting the raw emanation rate as it was obtained, the equilibrium number of radon atoms is also included. 
This is calculated using the half life of $^{222}$Rn to determine at what number of counts the balance between production and decay will be reached. 
The results are presented first on a `per fibre' basis and then adjusted to account for the total number of 110 fibres in the detector. 
The value in the last column is calculated considering that the emanated Radon atoms are fully dispersed in the total  outer water volume (5.3$\times$10$^6$ litres). 

The fibre mounting system plates and associated parts were not tested for Radon emanation, so we consider a "worst-case" scenario, in which all the Radon atoms produced in decays of the bulk $^{238}$U radioactivity are assumed to wander freely and disperse in the water volume. 
From the results shown in Table~\ref{tab:rad_results}, the conservative estimate yields a total rate per volume at equilibrium equivalent to 2\% of the guideline, well within the acceptance limit.

\begin{table}[t]
\begin{center}
	\begin{tabular} {| c | c | c || c |} \hline
    Material    & Counted Rate & Equilibrium Level & Rate per Volume \\ 
                & (atoms/day/fibre) & (atoms/fibre) & (mBq/m$^{3}$)\\ \hline
    fibres      & 53$\pm$33 & 290.6$\pm$181.0 & (1.21$\pm$0.75)$\times$10$^{-2}$\\

\hline
\hline

         & $^{238}$U Rate (mBq/kg) & Rate (mBq) & \\ \hline 
    Plates (PETG) & 6.22$\pm$4.83 & 82.10$\pm$63.75 & (1.55$\pm$1.20)$\times$10$^{-2}$\\
    Add. parts & 67.28$\pm$28.95 & 333.04$\pm$143.30 & (6.28$\pm$2.70)$\times$10$^{-2}$\\ \hline
    \hline
    \textbf{Total} &  &  & \textbf{(9.04$\pm$4.65)$\times$10$^{-2}$} \\ \hline

	\end{tabular}
\end{center}
\caption{Radon emanation budget for the PMT calibration system. 
The fibre results come from measurements of $^{222}$Rn emanation. 
The values for the other parts of the system come from a "worst-case" scenario estimate based on the bulk radioactivity results (Table~\protect\ref{tab:full_isotope_cont}). 
The total mass considered for the plates and additional materials was 16.0\,kg and 6.0\,kg, respectively, corresponding to 110 units. }
\label{tab:rad_results}
\end{table}

\subsection {Water compatibility}

The calibration system parts are located within the water volumes, both internal (mount plates) and external (fibres). 
The internal volume comprises the water between the acrylic vessel and the PMTs. 
The external water volume is also instrumented with PMTs and serves as a water Cherenkov veto detector for muons crossing SNO+. 
It is therefore important to prevent the installation of any materials that could leach optical impurities into the water. 
For this reason, all the materials of this calibration system  -- fibres, connectors, plates, auxiliary parts -- were screened for water leaching at the Brookhaven National Laboratory. 
After thorough cleaning, they were immersed in ultra-pure water, and the absorbance of this water was measured with an UV spectrometer. 
The first candidate material  for the fibre mount plates (see section~\ref{sec:mechanical}) -- black POM (polyoxymethylene) -- was clearly leaching into the water since the measured absorbance increased with immersion time, and so was replaced by white PETG.
For all the other tested materials the absorbance was found to be very low (below 0.002 for 10 cm cells) in the wavelength region of the PMT sensitivity and very stable over periods of several weeks to several months. 

Evaluation of the fibre degradation in the water is made both in laboratory measurements and in-situ. 
The chemical effects include a possible residual polymerisation that would modify the optical properties of the fibres, but as initial tests showed no degradation, this is not expected to be an issue. 
Dedicated accelerated ageing tests of the performance of the optical fibres in water over time are under way. 

The other control of the degradation is the installation of a monitoring channel, consisting of a fibre looped back onto the deck. 
One end of the fibre is inserted into an LED whose intensity is measured by a PIN diode (like all the other channels). Instead of being fixed on the PMT support structure, the detector-side end is fed back into a coupler (see Fig.~\ref{fig:LEDphotos}) with a PIN diode but no LED, allowing the output intensity to be measured as well. 
Regular runs of this fibre will provide the actual control of the possible ongoing degradation. 
The absence of the detector-side end connector in the water is the only change with respect to standard channels.

\section{Calibration system trigger}
\label{sec:control}

\begin{figure}[]
\begin{center}
\includegraphics[scale=0.45]{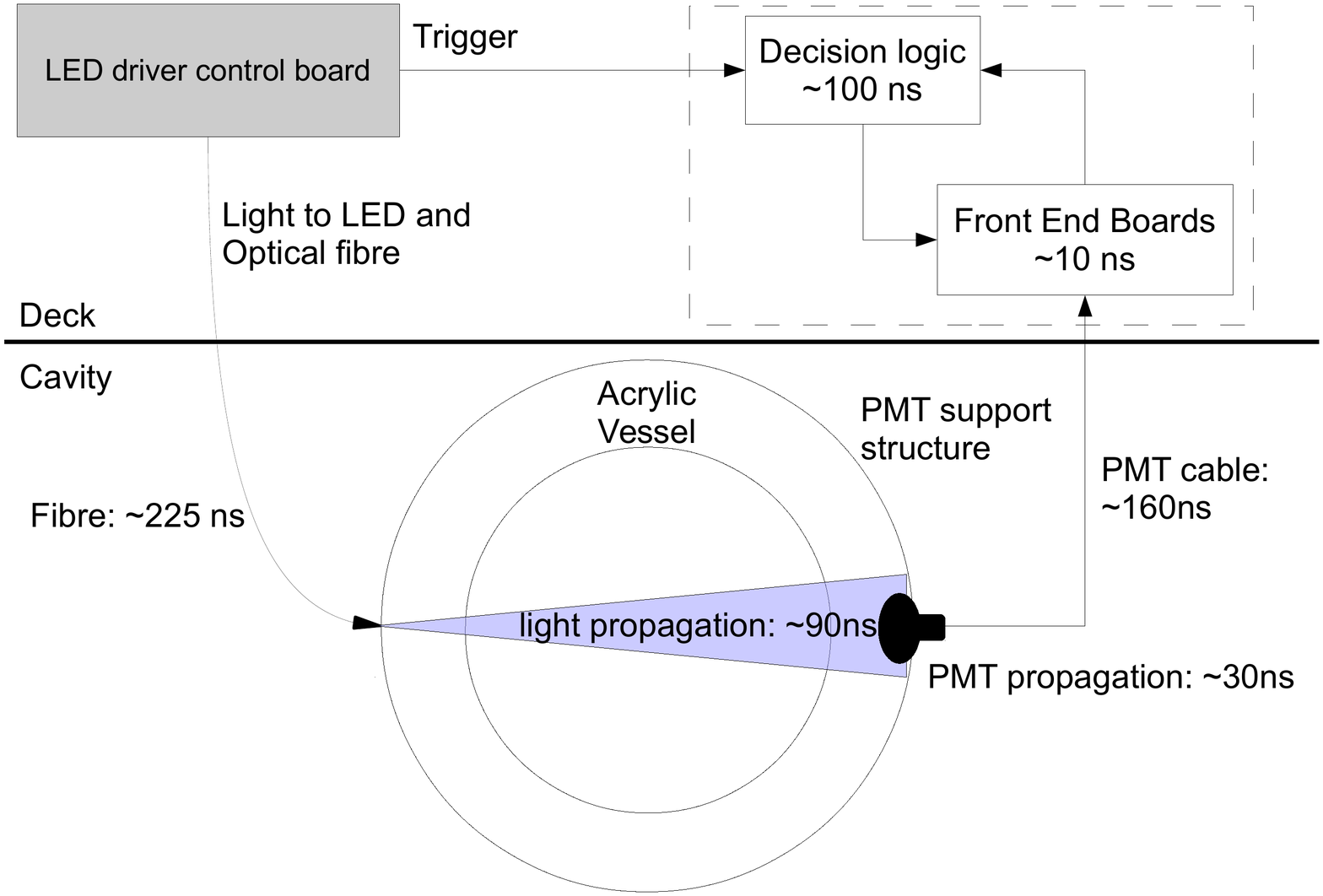}
\caption{Delays in a light calibration run. 
The control board must send out a light pulse and a trigger signal delayed by a predetermined time, to account for both the light propagation time in the fibre and in the detector on one hand, and in the PMTs and electronics on the other. 
The total light and signal time, from the emission to the acquisition of the data, is about 620\,ns.}
\label{fig:telliedelays}
\end{center}
\end{figure}

The SNO trigger and data acquisition systems, extensively described in~\cite{ref:SNOnim}, were upgraded to handle the higher counting rates expected in SNO+. 
The light calibration system will be connected into this existing setup so that it receives and acknowledges the detector trigger signal. 
This will allow the calibration to be synchronised with the SNO+ central clock and avoid the introduction of any new digital time jitter.

The integration of the light calibration control into the trigger system must let it retain its designed flexibility, i.e., the possibility to run various calibration sequences, while staying synchronised with the main trigger.  
Timing delays in the system, from light emission by the LEDs to the propagation in the fibres and PMT cables, and up to the trigger board's signal generation, must be accounted for.  
The sum of these delays is approximately 620\,ns (see schematic in Figure~\ref{fig:telliedelays}); the control board compensates for it through an adjustable delay of 0 -- 1275\,ns in steps of 5\,ns.  
In addition, each channel is equipped with an individual fine adjustment of 0 -- 63.75\,ns, in steps of 0.25\,ns.

The calibration system's electronics takes up one rack, located on the deck close to the feedthrough port where the fibres are routed into the cavity. 
The electronics comprises twelve boxes of LED drivers (each with eight LED drivers and one motherboard) and one control box.  
To monitor the calibration stability, each motherboard is equipped with five temperature sensors and each channel LED outputs light into a cone, described in section~\ref{sec:led}, that feeds both output fibre and PIN diode. 
The control box handles communications with the SNO+ digital master trigger board, sets signal amplitudes and delays for individual channels along with system-wide pulsing rates and signal/trigger delays, and extracts monitoring information from the PIN diodes and temperature sensors.

\begin{figure}[]
\begin{center}
\includegraphics[scale=0.45]{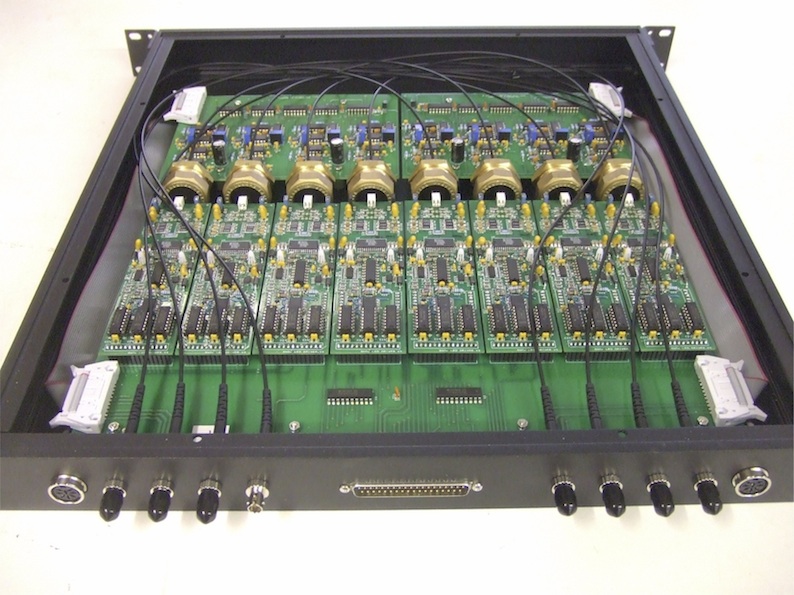}
\caption{One of the twelve boxes of the electronics of the calibration system. 
The eight boards in the front are the LED drivers. 
The LEDs and the PIN diodes are encapsulated in the  brass couplers and not visible. 
The boards in the back provide the intensity adjustments and read-out of the PIN diodes. 
The eight patch optical cables are fixed onto the back panel for connection with the LED-side end of the 45\,m fibres.}
\label{fig:elecbox}
\end{center}
\end{figure}

The available signal frequency from a single channel ranges from 3.9\,Hz up to 10\,kHz.  
Various calibration sequences are defined that include dedicated calibration runs at high frequency and calibration as part of standard physics runs, at Hz level.  
Standard runs consist of generating a defined number of signals from each LED in turn, with similar intensities from each LED, and the detector being force triggered for each signal. 
During calibration runs the temperature of the driver board is read and recorded at appropriate intervals.  
Once a given calibration sequence has been completed for an LED, the average pulse height measurement from the PIN diode is also recorded.

\section{Performance}
\label{sec:performance}
The goal of the performance study is to confirm that the light injection system is capable of producing the calibration pulses that are needed to characterise the time and charge response of each individual PMT. 
The relevant parameters that need to be validated are the intensity and angular coverage of the pulses, as well as their timing spread. 
This validation was carried out by performing realistic simulations of the beams from all positions, and by taking data with a partial set of the installed system, during commissioning runs with the air-filled detector.
This section gives an overview of the full detector simulations that have been performed and the comparison of these simulations to commissioning data.
\begin{figure}[htbp]
\centering
\includegraphics[width=0.98\textwidth]{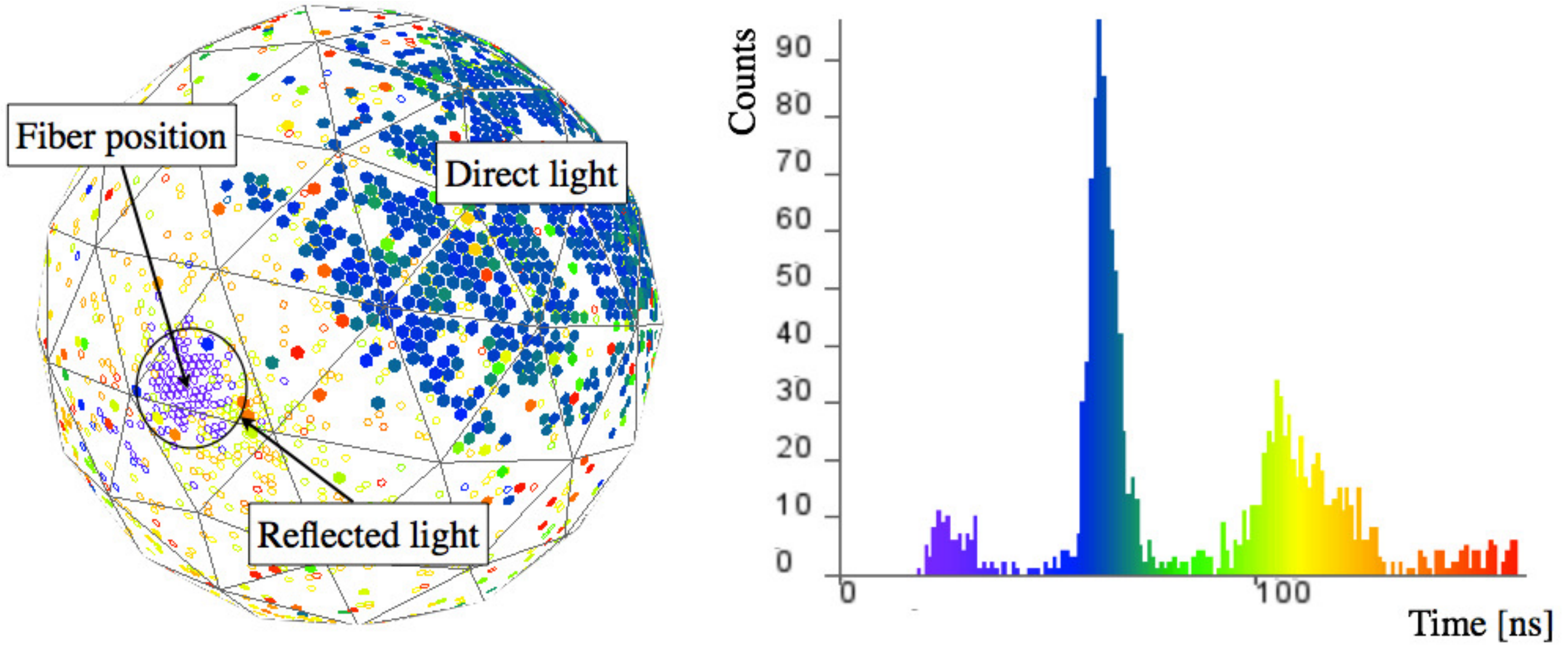}
\caption{3D viewer display (left) for a simulated high intensity light injection event inside the SNO+ detector. 
Each dot represents a PMT, and open circles represent PMTs on the far side of the detector as viewed. 
The color indicates the time at which the PMT triggered, according to the time histogram (right). }
\label{fig:eventSim}
\end{figure}

\subsection{Simulation of light pulses}

The simulation and analysis tool of the SNO+ experiment makes use of GEANT4~\cite{ref:geant} and ROOT~\cite{ref:root} libraries and has been verified to accurately model the detector using the well-calibrated simulation of the SNO experiment. 
This framework provides a complete simulation of particle generation and transport in the detector, and of the production, propagation and detection of optical photons. 
Optical processes in the detector media are included, such as absorption and scattering in the scintillator, reflection and refraction using a detailed detector geometry, as well as the simulation of the PMT-response including the single-photon charge spectra and the full DAQ-chain. 

An LED/fibre beam event generator was developed and included in the simulation, allowing the generation of light pulses for each selected fibre mounting position, and the individual setting of the beam intensity, spectrum and angular emission profile. 
The simulations presented in this section were produced using the ex-situ measurements of wavelength, angular distribution and time spread described in Sections \ref{sec:led} and \ref{sec:fibres} as input. 
In this way, the performance of the light injection calibration system can be investigated and the calibration analysis code that will perform the hit-level time and charge PMT calibration can be tested. 

Figure~\ref{fig:eventSim} shows a representation of a simulated high-intensity event inside the SNO+ detector. 
The time histogram shows the early hits in purple. This is light that is reflected off the acrylic vessel and detected by the PMTs that are located around the LED position. 
The forward light, in blue, has a larger distance to travel to the PMTs that are located on the opposite side of the detector. 
Light that is reflected back from those PMTs makes up the yellow population of hits. 
Scattered light will arrive still later and is shown in orange.

\subsection{Expected performance with full system}
As was seen in Section~\ref{sec:led}, the intensity of each LED can be adjusted individually. 
The dynamic range of each LED/fibre combination must obey the following requirements. 
If the intensity is too low, the time it will take to accumulate the necessary statistics for the PMT calibration becomes prohibitively long, especially in the high charge tail region. 
On the other hand, too many photons in a single pulse will translate into a higher percentage of multi-photon hits at the target PMTs and therefore into contaminated single-photon charge spectra. 
The goal of the simulations is to determine the optimal intensity in order to reach these requirements.

A check of the number of multi-photon hits as a function of intensity was performed for different phases of the SNO+ experiment. 
In a first phase, the acrylic vessel will be filled with water. 
This water will then be replaced by the liquid scintillator. 
Finally, Tellurium will be added to the liquid scintillator for the neutrino-less double beta decay phase.  
The optical properties of water, loaded and unloaded scintillator are different, and therefore the transmission of light through the acrylic vessel volume changes between the different phases. 

The dynamic range of the pulses must be such that enough light gets through to the PMTs, while also keeping the multi-photoelectron hit contamination below 1$\%$. 
Using the Poisson distribution, that corresponds to a maximum hit probability of around 5$\%$. 
The LED pulse intensity can be set for the different SNO+ phases, in order for the hit probability to be close to the maximum of 5$\%$ for the PMTs that receive the highest intensity from the fibre, the ones within the central region, 14.5$^{\circ}$ around the fibre pointing direction. 
From simulations, it was found that the required intensity for the single-photon charge calibration lies between 10$^{2}$ to 10$^{4}$ photons per pulse at the detector-side end of the fibre for the water phase and the loaded scintillator phase, respectively.

With this setting, the expected hit probability in the border region, between 14.5$^{\circ}$ and 29$^{\circ}$ around the fibre pointing direction, is about 1$\%$.
This value is then used to calculate the expected total time needed to collect a complete calibration data-set, that depends as well on the desired time resolution and the frequency at which the  pulses are fired. 

 Assuming a required time precision of 0.5\,ns, a pulse width of 5.2\,ns and a 10\,Hz pulse frequency, this gives a total required run duration of approximately 5~hours to accumulate enough data for the border region. 
 This is the duration required when running at low rate within a Physics run. 
 Dedicated calibration runs can be taken at least at 1\,kHz and will thus be feasible in only a few minutes. 
 With this performance a flexible calibration program, with fast dedicated runs and low rate continuous running will be possible in SNO+.

For charge spectrum studies and calibration, it is enough that each PMT is hit by light from at least one fibre. 
However, for the PMT time calibration, a more redundant data-set is required, where a significant fraction of the PMTs detect light from more than one fibre. 
This redundancy will allow for the time offsets between the different LEDs to be extracted and corrected for in the PMT time calibration. 

In terms of PMT coverage, the first level of redundancy in the system is ensured by having two fibres in each node. 
The fibre positions in the 91 PMT structure nodes have been chosen in such a way that all PMTs will see direct light from at least one fibre node, and in most cases, from two nodes.  
The two-node redundancy requirement faces two difficulties. 
The neck of the AV determines a "hole" in the PMT support structure, effectively removing its 92$^{nd}$ node and reducing the coverage of the PMTs in the bottom of the PSUP. 
In order to circumvent this limitation, two extra fibre cables are positioned in non-node locations close to the AV neck. 
In addition, the hold-down and hold-up rope systems that surround the AV cause shadows to the fibre beams. 
In several cases, this shadow is large enough to block or significantly reduce the intensity seen in the opposite PMTs.   

Simulations of the full system, that include a detailed model of the rope system, show that even with those shadows, most PMTs receive light from more than five different fibres. 
Considering 10\% of the central region intensity as the acceptance limit for worse illuminated PMTs, the simulations show that all PMTs see at least one fibre node, and only 20 see only one fibre node.
If that acceptance is relaxed to 2\% of the central region intensity, then all PMTs are illuminated by at least two fibre nodes. 
This means that, in the unlikely event of failure of the two fibres that illuminate those 20 PMTs, they can still be calibrated, with dedicated runs of a neighbouring fibre, at higher intensity.

Equally important, all fibres have enough overlap so that any relative delay can be measured. 
This is verified with the PMT calibration code, which extracts time offsets between individual fibre pairs by combining the data from the overlapping PMTs. 
A minimum of 20 PMTs need to see hits from the same fibre pair before the time offset between those two fibres is extracted. 
Simulations also show that the number of PMTs that see each adjacent fibre pair\,\footnote{i.e., fibres for which the detector-side end is within 4.5\,m of each other, or 30$^{\circ}$ measured from the detector center.} is always above 30 and around 80 for most pairs.

\begin{figure}[ht]
        \centering
                \includegraphics[width=0.49\textwidth]{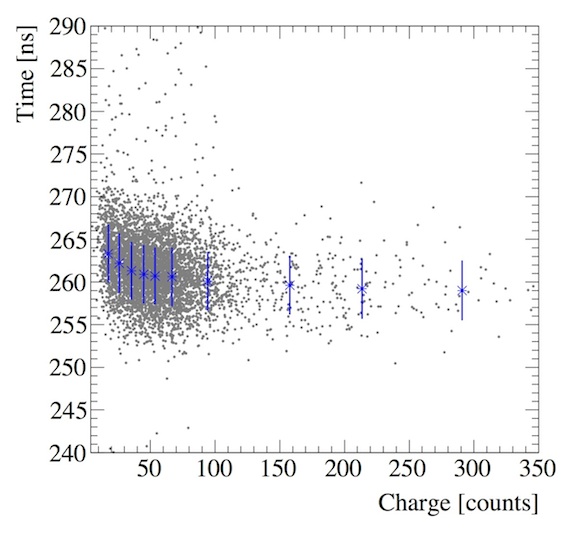}
                \includegraphics[width=0.49\textwidth]{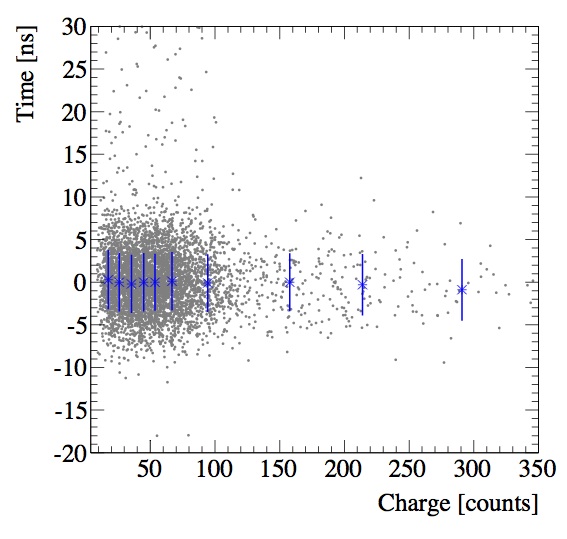}

        \caption{Left: An example of the discriminator walk effect in simulated light injection data. 
        The data points with error bars correspond to an average evaluated in a given charge bin and used to interpolate the time-charge dependence. 
        Right: simulated light injection data after applying the extracted time corrections. For a comparison with SNO data, see Fig. 3.2 of reference~\protect\cite{ref:cameron}.}
\label{fig:PCAProof}
\end{figure}

The main goal of the PMT timing calibration algorithm is to provide hit-time corrections to account for each channel's offset and for the dependence of time on charge, i.e. the fixed discriminator effect (see Figure~\ref{fig:PCAProof} (Left)). 
Several million light injection events were simulated and used to extract timing calibration constants which were then applied to calibrate simulated data of uniform light pulses emitted in the center of the SNO+ detector. 
For the simulated detector, the calibration performed with the light injection system has been shown to effectively correct for the charge dependency of the time response, see Figure~\ref{fig:PCAProof}(Right).

\subsection{Commissioning with partial system}
Commissioning periods of the air-filled SNO+ detector took place in October 2012, March 2013 and February 2014.  
During these runs, 36 of the 110 light injection fibres were installed on the mounting points of the lower hemisphere of the PSUP. 
For the first and the last run, all the SNO+ electronic crates were powered on and the majority of the PMTs were supplied with high voltage. 
Seven of the nineteen crates were on for the second run. 
The light injection event intensity was high (around 10$^{5}$ photons per pulse) for the first run. 
The second and third run had a flexible intensity (from 10$^{3}$ to 10$^{5}$ photons per pulse). 
Using the commissioning data, two verifications for the light injection system were performed: coverage and charge response. 

\begin{figure}[h]
        \centering
                \includegraphics[width=0.6\textwidth]{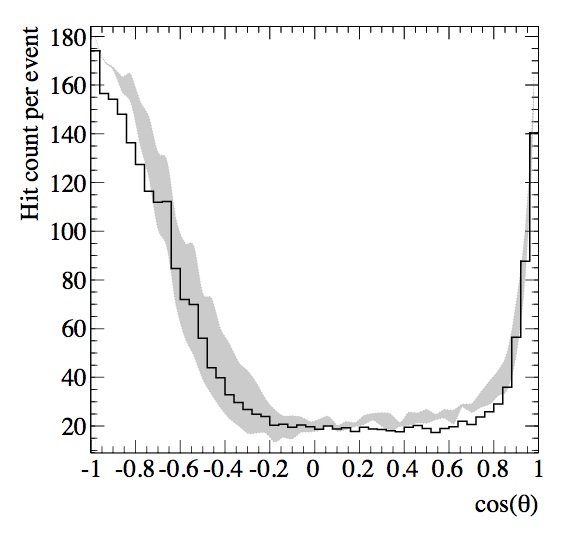}
        \caption{Number of hits per event versus the cosine of the angle between PMT and fibre position for commissioning data (black, thick solid). 
        The light grey band represents the region between minimum (simulation with 11\,$^{\circ}$ opening angle and 9$\times$10$^{4}$ photons per pulse) and maximum (13 degree opening angle and 1.1$\times$10$^{5}$  photons per pulse) hit probability. 
        The direct and reflected light are detected at cosine -1 and 1 respectively. }
        \label{fig:led71-hitprob}
\end{figure}

\begin{figure}[h]
        \centering
                \includegraphics[width=0.6\textwidth]{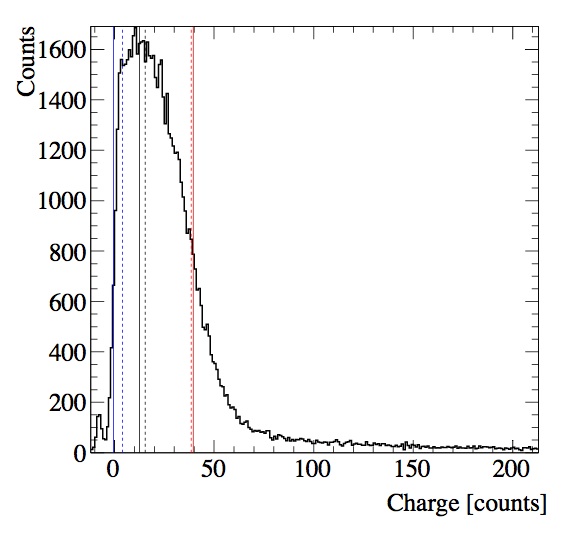}
        \caption{ An example of the PMT charge calibration constant extraction using low-intensity light injection data. 
        The dashed (solid) lines show the position of the threshold, peak and high-half point from left to right for SNO (SNO+) data.}
                \label{fig:chargecal}
\end{figure}

First, all installed fibres were tested and their positions and pointing directions were compared to the values that were inserted in the database after fibre installation. All fibres were found to be undamaged and with correctly matching positions and directions. 
Figure~\ref{fig:led71-hitprob} shows the number of hit PMTs per event versus the cosine of the angle between PMT and fibre with respect to the center of the detector for data and two simulations of the same fibre, with different angular distributions.

Since the intensity is quite high, the comparison of the measured number of detected hits at the PMTs with simulations is expected to be distorted by MPE effects in the forward region. 
From Figure~\ref{fig:2dscan}, an opening angle of  11\,$^{\circ}$ to 13\,$^{\circ}$ is expected in air. 
The data at high intensity had between 9$\times$10$^{4}$ and  1.1$\times$10$^{5}$ photons per event. 
From this, a minimum and maximum hit probability was simulated, represented by the light grey area in Figure~\ref{fig:led71-hitprob}, showing a good agreement with data.

Second, the PMT charge calibration code used the low-intensity commissioning data to verify that the single-photon-electron level was achieved. 
To characterise the charge spectrum, the PMT calibration analysis extracts three parameters for each PMT: the threshold, peak and high-half point. 
The latter one is then used in simulations to scale the charge spectrum for each individual PMT. 
Figure~\ref{fig:chargecal} shows the charge spectrum for a single PMT with the extracted values for the three spectrum parameters, compared to the values that were extracted for SNO using a low-intensity central light source. 
The small differences seen between SNO and SNO+ are likely caused by a difference in the electronics calibration.
Since the light attenuation is expected to increase when the detector is filled, the data taken during the commissioning of the SNO+ light injection system shows that the pulse intensity can be lowered to a level where the multi-photon hit contamination is minimal. This means that the light injection system will be able to produce events that are low enough in intensity to be used for the PMT charge calibration.

\section{Conclusions}

A non-invasive system for the calibration and monitoring of the SNO+ experiment's array of 9500 photomultipliers was designed, installed and tested. 
The system is based on the use of optical fibres to inject light pulses produced externally by LEDs and their dedicated fast electronics drivers. 

This design presents several advantages. 
The use of a permanently installed light transmission system external to the detector's target volume is crucial for low background experiments, such as those dedicated to neutrinoless double-beta decay or solar neutrinos, since it allows regular monitoring with no risk of radon ingress. 
The concept of using light beams that cross the whole detector and illuminate PMTs on the opposite side allows for a simple system, with a small number of calibration channels. 
In the SNO+ implementation, about 100 PMTs can be calibrated with each fibre/LED channel. 
This requires a bootstrapping method for the data analysis, and makes some degradation of the time width - due to scattering - unavoidable. 
However, that can be compensated by higher statistics in the calibration data sample. 

The design and quality control tests for this system were presented. 
The wavelength distribution was found to be very uniform, with a spread on the peak measurement of only 2.6\,nm (sigma) across all the final 96 LEDs. 
The overall effective light yield of the LED/fibre system has a large spread, varying by more than an order of magnitude, due mostly to the LED-fiber coupling.
However, the driver pulse current is individually adjustable for each LED and can compensate that. 
All channels are able to cover the whole range of required intensities from $10^3$ to $10^5$ photons.
The timing performance of the LED/fibre combination was also found to match the requirements, with a FWHM of 4.4\,ns to 5.2\,ns over the operational range of intensities. 
The spread of rise time measured over all fibres during quality control, was~0.1\,ns, so the uniformity is very good.

As for the angular distribution, three measurements were carried out: a detailed 2D scan of a single fibre, a semi-automated 1D measurement for all 220 fibres, and a measurement in the detector itself, during the air commissioning phase. 
They all gave compatible results, and showed that the fibre beam is wide enough to allow the required PMT coverage. 
These commissioning test results were combined with extensive Monte Carlo simulations in order to prove that the system is expected to perform well and calibrate all the SNO+ PMTs to the required accuracy.

\section{Acknowledgements}

We would like to thank the SNO+ Collaboration and the SNOLAB technical staff for their fundamental and continued support for this work at various stages, from discussions on the design to installation and commissioning work at site.
In particular, we are very grateful for the crucial work of B. Cleveland and I. Lawson on the radiopurity screening of the various materials, to S. Hans and M. Yeh for the leaching tests, 
and to K. Singh for discussions on the commissioning data analysis.
We would also like to thank the technical staff at the electronics workshops of the Universities of Sussex and Leeds and at the LIP mechanical workshop.

This research was supported in part by: national funds from Portugal and by European Union FEDER funds through the COMPETE program,  through FCT (Funda\c{c}\~{a}o para a Ci\^{e}ncia e a Tecnologia) with the project grant PTDC/FIS/115281/2009; the Science and Technology Facilities Council (STFC), United Kingdom, through grants ST/J001007/1 and ST/K001329/1, the European Union's Seventh Framework Programme FP7/2007-2013, under the European Research Council (ERC) grant agreement 278310 and the Marie Curie grant agreement PIEF-GA-2009-253701; the University of Leeds; the Canada Foundation for Innovation and the Natural Sciences and Engineering Research Council of Canada. This material is based upon work supported by the Director, Office of Science, of the U.S. Department of Energy under Contract No. DE-AC02-05CH11231; the U.S. Department of Energy, Office of Science, Office of Nuclear Physics, under Award Number DE-SC0010407; the National Science Foundation under Grant No. NSF-PHY-1242509; and the University of 
California, Berkeley.

\bibliographystyle{model1-num-names}

\begin{thebibliography}{20}


\bibitem{ref:SNOnim} J. Boger \emph{et al.}, Nucl. Instr. and Meth. A 449 (2000) 172-207.
\bibitem{ref:snoplus_proc} M. Chen, Nucl. Phys. B (Proc. Suppl.) 154 (2005) 65-68.
\bibitem{ref:snoplus_proc2} V. Lozza, RPSCINT 2013, EPJ Web of Conferences Vol. 65 (2014).
\bibitem{ref:ropenet} A. Bialek \emph{et al.} (in preparation).
\bibitem{ref:SNEWS} P. Antonioli \emph{et al.}, New J. Phys. 6 (2004) 114.
\bibitem{ref:laserball} B.A. Moffat \emph{et al.}, Nucl. Instr. and Meth. A 544 (2005) 255-265.
\bibitem{ref:borexino} G. Alimonti \emph{et al.}, Nucl. Instr. and Meth. A 600 (2009) 568-593.
\bibitem{ref:borexino_fibres} B. Caccianiga \emph{et al.}, Nucl. Instr. and Meth. A 496 (2003) 353-361.

\bibitem{ref:cameron} J.~Cameron, \textit{The photomultiplier tube calibration of the Sudbury Neutrino Observatory}, Ph.D. Thesis, Oxford University (2001).
\bibitem{ref:PMT_test} C.J.~Jillings \emph{et al.}, Nucl. Instr. and Meth. A 373 (1996) 421-429.

\bibitem{ref:LED} E. Fred Schubert, Light-Emitting Diodes, $2^{nd}~edition$, Cambridge University Press, 2010.
\bibitem{ref:LEDspecs} Brite-LED Optoelectronics, Ultra Brightness Cyan LED lamp, BLÐLUCY5N15C series. 
Web: \url{http://www.brite-led.com/PDF/BL-LUCY5N15C series datasheet.pdf}   [Accessed April 19, 2014].
\bibitem{ref:spectrometer} Ocean Optics model Maya2000 Pro.

\bibitem{ref:circuit} Differential ultra-fast laser diode driver circuit IXLDO2SI from the IXYS company (USA).

\bibitem{ref:pmt} Hamamatsu H10721P-110 PMT.
\bibitem{ref:singlePE} J. E. McMillan, The Single Photon Technique for Measuring LED Pulser Flash Width, Sheffield Particle Astrophysics (Sheffield: Department of Physics and Astronomy Internal Publication, Sheffield University)(1999).
\bibitem{ref:scope} Tektronix DPO 3054 Oscilloscope.
\bibitem{ref:powermeter} ThorLabs PM100USB light-meter.


\bibitem{ref:fibre} Mitsubishi Super Eska SH4002, Polyethylene jacketed, PMMA optical fibre cord. Web: \url{http://i-fiberoptics.com/fiber-detail.php?id=15&sum=80} [Accessed February 18, 2014].
\bibitem{ref:fibrometer} J.~Silva \emph{et al.}, Nucl. Instr. and Meth. A 580 (2007) 318-321;\\ M. David {\it et al}, ATLAS public note ATL-TILECAL-PUB-2008-003, (2007).

\bibitem{ref:gamma_assay} I. Lawson, B. Cleveland, AIP Conf. Proc. 1388, 68 (2011).
\bibitem{ref:counting_results} http://www.snolab.ca/users/services/gamma-assay/HPGe\_samples\_snoplus\_master.html [Accessed: 18 February 2014].
\bibitem{ref:sno_radon} I.Blevis \emph{et al.},  Nucl. Instr. and Meth. A 517 (2004) 139-153.
\bibitem{ref:radon_queens} M. Seddighin, {\textit Low energy $^8$B solar neutrinos in SNO+: Controlling and Constraining Radon Backgrounds}, M.Sc. Thesis, Queen's University (2013).

\bibitem{ref:geant} S. Agostinelli \emph{et al.}, Nucl. Instr. and Meth. A 506 (2003) 250-303.
\bibitem{ref:root} I. Antcheva  \emph{et al.}, Comput. Phys. Comm. 180 (2009) 2499-2512.
\end{thebibliography}

 
\end{document}